\newcommand{\aj}{Astron. J.}
\newcommand{\mnras}{Mon. Not. R. Astron. Soc.}
\newcommand{\jcap}{J. Cosmol. Astropart. Phys.}
\newcommand{\physrep}{Phys. Rep.}
\newcommand{\aap}{Astron. Astrophys.}
\newcommand{\apjl}{Astrophys. J. Lett.}
\newcommand{\pasa}{PASA}

\documentclass[twocolumn,aps,prd,nofootinbib,showpacs,superscriptaddress]{revtex4-1}
\usepackage{amsmath,graphicx,bm,color}
\begin{document}
\title{Eliminating the optical depth nuisance from the CMB with $21\,\textrm{cm}$ cosmology}

\author{Adrian Liu}
\email{acliu@berkeley.edu}
\email{Hubble Fellow}
\affiliation{Department of Astronomy, UC Berkeley, Berkeley, CA 94720, USA}
\affiliation{Berkeley Center for Cosmological Physics, UC Berkeley, Berkeley, CA 94720, USA}

\author{Jonathan R. Pritchard}
\affiliation{Imperial Center for Inference and Cosmology, Imperial College London, Blackett Laboratory, Prince Consort Road, London SW7 2AZ, United Kingdom}

\author{Rupert Allison}
\affiliation{Sub-department of Astrophysics, University of Oxford, Denys Wilkinson Building, Oxford, OX1 3RH, United Kingdom}

\author{Aaron R. Parsons}
\affiliation{Department of Astronomy, UC Berkeley, Berkeley, CA 94720, USA}
\affiliation{Radio Astronomy Laboratory, UC Berkeley, Berkeley, CA 94720, USA}

\author{Uro\v{s} Seljak}
\affiliation{Department of Astronomy, UC Berkeley, Berkeley, CA 94720, USA}
\affiliation{Berkeley Center for Cosmological Physics, UC Berkeley, Berkeley, CA 94720, USA}
\affiliation{Department of Physics, UC Berkeley and Lawrence Berkeley National Laboratory, Berkeley, CA 94720, USA}

\author{Blake D. Sherwin}
\affiliation{Berkeley Center for Cosmological Physics, UC Berkeley, Berkeley, CA 94720, USA}
\affiliation{Department of Physics, UC Berkeley and Lawrence Berkeley National Laboratory, Berkeley, CA 94720, USA}
\affiliation{Miller Institute for Basic Research in Science, University of California, Berkeley, CA, 94720, USA}

\date{\today}


\begin{abstract}
Amongst standard model parameters that are constrained by cosmic microwave background (CMB) observations, the optical depth $\tau$ stands out as a nuisance parameter. While $\tau$ provides some crude limits on reionization, it also degrades constraints on other cosmological parameters. Here we explore how $21\,\textrm{cm}$ cosmology---as a direct probe of reionization---can be used to independently predict $\tau$ in an effort to improve CMB parameter constraints. We develop two complementary schemes for doing so. The first uses $21\,\textrm{cm}$ power spectrum observations in conjunction with semi-analytic simulations to predict $\tau$. The other uses global $21\,\textrm{cm}$ measurements to directly constrain low redshift (post-reheating) contributions to $\tau$ in a relatively model-independent way. Forecasting the performance of the upcoming Hydrogen Epoch of Reionization Array, we find that significant reductions in the errors on $\tau$ can be achieved. These results are particularly effective at breaking the CMB degeneracy between $\tau$ and the amplitude of the primordial fluctuation spectrum $A_s$, with errors on $\ln (10^{10} A_s)$ reduced by up to a factor of four. Stage 4 CMB constraints on the neutrino mass sum are also improved, with errors potentially reduced to $12\,\textrm{meV}$ regardless of whether CMB experiments can precisely measure the reionization bump in polarization power spectra. Observations of the $21\,\textrm{cm}$ line are therefore capable of improving not only our understanding of reionization astrophysics, but also of cosmology in general.
\end{abstract}

\maketitle

\section{Introduction}
\label{sec:Intro}
Through a complementary blend of cosmological probes, the last decade has seen the emergence and strengthening of a concordance $\Lambda$CDM model of our Universe. Using just a handful of parameters, the $\Lambda$CDM model provides an adequate fit to data from a wide range of epochs in our cosmic timeline, ranging from Big Bang Nucleosynthesis (BBN) to the Cosmic Microwave Background (CMB) to galaxy surveys and supernovae measurements.

Examined in more detail, however, tensions have emerged between various datasets. Consider the latest CMB results from the \emph{Planck} satellite \cite{Planck2015overview}, for instance. Distance measures inferred from \emph{Planck} are in mild tension with Lyman-$\alpha$ baryon acoustic oscillation (BAO) constraints derived from quasar observations \cite{Planck2015parameters}. As another example, \emph{Planck} data is best fit by a higher amplitude of density fluctuations than is preferred by measurements of weak lensing and galaxy cluster counts \cite{Planck2015clusters}. While currently still tolerable, these tensions may be the result of experimental systematics, or may be the first sign of new physics.

To make progress, it will be necessary to sharpen our cosmological constraints. In doing so, the hint of inconsistencies between data sets will either vanish or become statistically significant. One way to accomplish this is to simply take more data. Galaxy surveys, for instance, are poised to significantly improve their reach with new experiments such as the Dark Energy Spectroscopic Instrument (DESI) \cite{levi_et_al2013}. With the CMB, on the other hand, it is likely that many improvements will come from exploiting qualitatively new probes, such as a measurement of the primordial B-mode signal, or better measurements of CMB lensing and secondary anisotropies. These have the ability to access previously unconstrained phenomena, as well as to break existing degeneracies between cosmological parameters. Better measurements will also pave the way for expanded cosmological models that constrain the neutrino mass or the time-evolution of dark energy.

In this paper, we examine the role that the emerging field of $21\,\textrm{cm}$ cosmology can play in sharpening CMB constraints. With $21\,\textrm{cm}$ cosmology, one seeks to use the $21\,\textrm{cm}$ hyperfine transition to map the large scale distribution of neutral hydrogen at a variety of redshifts. Existing and upcoming efforts include lower redshift efforts ($z \lesssim 2$) to target baryon acoustic oscillations as well as higher redshift measurements that will provide a uniquely \emph{direct} probe of the intergalactic medium (IGM) during the reionization epoch, when radiation from the first galaxies systematically ionized the IGM. Examples of $21\,\textrm{cm}$ experiments include the Green Bank Telescope \cite{masui_et_al2013,switzer_et_al2013}, the Canadian Hydrogen Intensity Mapping Experiment \cite{shaw_et_al2014a}, and the Baryon Acoustic Oscillation Broadband and Broad-beam Array \cite{pober_et_al2013} at low redshifts, and the Precision Array for Probing the Epoch of Reionization (PAPER \cite{parsons_et_al2010}), the Murchison Widefield Array (MWA \cite{tingay_et_al2013,bowman_et_al2012}), the Giant Metrewave Radio Telescope \cite{paciga_et_al2013}, and the Low Frequency Array (LOFAR \cite{van_haarlem_et_al2013}) at high redshifts. Although the methods in this paper are general, we will focus on the upcoming high-redshift experiment Hydrogen Epoch of Reionization Array (HERA \cite{pober_et_al2014}) as our main worked example, with some quick estimates also provided for the future Square Kilometre Array (SKA \cite{mellema_et_al2013}).

Whereas reionization is the prime epoch of study for many $21\,\textrm{cm}$ experiments, it is simultaneously an interesting epoch and a nuisance for CMB studies. Reionization releases free electrons into the IGM, which Compton scatter CMB photons as they stream from the surface of last scattering to our detectors, necessitating the introduction of an optical depth parameter $\tau$ that quantifies the probability of scattering. The free electrons released by reionization source additional polarization fluctuations, giving rise to a ``reionization bump" feature at large angular scales in polarization power spectra. The amplitude of the bump scales as $\tau^2$, and thus an accurate measurement of this feature enables precise constraints on $\tau$. In turn, $\tau$ can be converted into a crude redshift of reionization, with higher $\tau$ implying a higher redshift.

At fine angular scales, the main effect of $\tau$ is to dampen the measured CMB anisotropies (whether in the temperature or polarization power spectra), which unfortunately means that $\tau$ is largely degenerate with $A_s$, the amplitude of primordial density fluctuations. Although this degeneracy is partially broken by the aforementioned polarization signature (or by CMB lensing if one assumes no departures from standard $\Lambda$CDM evolution), it remains to a large extent. This degrades cosmological parameter constraints from the CMB, and it is in this sense that reionization is a nuisance for CMB experiments.

In this paper, we show that $21\,\textrm{cm}$ reionization experiments have the ability to place constraints on reionization that are stringent enough to allow high-precision determinations of $\tau$. These can then be fed into CMB studies, effectively eliminating $\tau$ as a nuisance parameter. Provided astrophysical modeling uncertainties are made sufficiently small with upcoming measurements, this would push CMB measurements into a new regime by avoiding cosmic variance limits on a determination of $\tau$. Concretely, a known value of $\tau$ would improve estimates of $A_s$. In turn, this would sharpen any cosmological tests that depend on comparing primordial fluctuations (controlled by $A_s$) and low-redshift measures of structure such as cluster counts and CMB lensing. Any discrepancies between early and late time measurements are potentially indicative of cosmological evolution beyond that predicted by basic $\Lambda$CDM, signaling cosmological evidence for model extensions such as a non-zero neutrino mass or an evolving dark energy equation of state. Measurements of the $21\,\textrm{cm}$ line will therefore have broad cosmological implications for future CMB studies.

This work differs from previous cosmological parameter estimation forecasts in that previous papers have mostly arrived at improved constraints by focusing on the larger co-moving volume of our Universe that can be potentially accessed by $21\,\textrm{cm}$ surveys compared to traditional galaxy surveys \cite{mcquinn_et_al2006,bowman_et_al2007,mao_et_al2008,loeb_and_wyithe2008,tegmark_and_zaldarriaga2009,barger_et_al2009,visbal_et_al2009,clesse_et_al2012,oyama_et_al2013,bull_et_al2015}. The framework that we establish here assigns a more limited---but arguably more robust---role to $21\,\textrm{cm}$ surveys. In this paper, the CMB experiments deliver the bulk of the cosmological information, and the $21\,\textrm{cm}$ surveys play the secondary role of providing details about reionization that are difficult to obtain from the CMB. In this sense, our work builds on that of Ref. \cite{pritchard_et_al2010}, where the possibility of estimating $\tau$ from $21\,\textrm{cm}$ cosmology was briefly considered. Ref. \cite{clesse_et_al2012} also emphasized the self-consistency between CMB and 21cm that is crucial to the current work. Our approach here is complementary to theirs in that they rely on phenomenological fits to numerical simulations with tunable nuisance parameters, whereas we ascribe a more central role to the detailed astrophysics of reionization as modeled by semi-analytic codes. The rest of this paper is organized as follows. In Sec. \ref{sec:fidExpt} we introduce the fiducial experiments and models that we use for our forecasts. Sec. \ref{sec:Ingredients} discusses the various sources of uncertainty in a prediction of $\tau$. Sec. \ref{eq:Pkformalism} then establishes a formalism for folding $21\,\textrm{cm}$ power spectrum measurements into CMB analyses via $\tau$. Forecasted improvements on cosmological parameters based on this formalism are presented in Sec. \ref{sec:CMBresults}. In Sec. \ref{sec:GlobalSig} we explore how direct measurements of the mean $21\,\textrm{cm}$ brightness temperature field can reduce the model dependence of a $\tau$ prediction, and we summarize our conclusions in Sec. \ref{sec:conc}.

\section{Fiducial experiments and assumptions}
\label{sec:fidExpt}

Throughout this paper, we will illustrate our framework for sharpening cosmological constraints by considering various fiducial experiments. From the $21\,\textrm{cm}$ side, we will consider two types of experiments. Sections \ref{eq:Pkformalism} and \ref{sec:CMBresults} concentrate on $21\,\textrm{cm}$ power spectrum experiments. These typically consist of low-frequency radio interferometers, which measure the redshifted brightness temperature contrast $\delta T_b (\mathbf{\hat{n}}, \nu)$ of the $21\,\textrm{cm}$ line against the CMB, where $\mathbf{\hat{n}}$ specifies the direction on the sky and $\nu$ is the observation frequency. Given the spectral nature of the probe, different frequencies can be translated into different radial distances, and the result is a three-dimensional brightness temperature distribution $\delta T_b (\mathbf{r})$ in terms of comoving coordinates $\mathbf{r}$. Fourier transforming and binning this distribution then allows a measurement of the brightness temperature power spectrum $P_\textrm{21} (k)$, defined by
\begin{equation}
\label{eq:PspecDef}
\langle \delta \widetilde{T}_b(\mathbf{k}) \delta \widetilde{T}_b(\mathbf{k}^\prime)^* \rangle \equiv (2\pi)^3 \delta^D (\mathbf{k} - \mathbf{k}^\prime ) P_\textrm{21} (k),
\end{equation}
where $\delta^D$ signifies a Dirac delta function, pointed brackets $\langle \cdots \rangle$ represent an ensemble average, and $ \delta \widetilde{T}_b(\mathbf{k})$ is the Fourier transform of  $\delta T_b$ evaluated at spatial wavevector $\mathbf{k}$. Note that since the statistical properties of the brightness temperature evolve substantially as reionization progresses, one typically does not form a single power spectrum over the entire survey volume of a $21\,\textrm{cm}$ survey, as doing so would violate the central assumption of translation invariance necessary for forming a power spectrum. Instead, most analysts break up their wide bandwidth data into a few (relatively) narrow chunks and compute multiple power spectra $P_\textrm{21} (k,z)$ centered on several different redshifts. Current instruments such as GMRT, MWA, LOFAR, and PAPER have begun to place scientifically interesting upper limits on such power spectra \cite{parsons_et_al2014,ali_et_al2015,pober_et_al2015,greig_et_al2015a}.

As our fiducial $21\,\textrm{cm}$ power spectrum experiment, we pick HERA, a low-frequency radio interferometric array that is currently being constructed in the South African Karoo desert. HERA's construction plans involve an incremental buildup of a series of $14$-m diameter dishes, closely packed in a hexagonal configuration. In this paper, we assume that observations are made when the array consists of $331$ such dishes. These dishes are not steerable, and instead observe in a drift-scan mode. From HERA's location, this provides roughly $6\,\textrm{hours}$ of usable observation time per day, defined to be when the Galactic plane is sufficiently far below the horizon. We further assume $180\,\textrm{days}$ of observations, providing $1080\,\textrm{hours}$ of total observation time. This should, however, not be considered $1080\,\textrm{hours}$ of integration time in the conventional sense, since drift-scan observations are by definition distributed amongst different patches of the sky. The observation time is thus only coherently integrated for a portion of the this time, although all the data are eventually folded into a single final estimate of the power spectrum.

To forecast power spectrum sensitivities amidst such complications, we make use of the {\tt 21cmSense} code\footnote{{\tt https://github.com/jpober/21cmSense}} \cite{pober_et_al2013,pober_et_al2014}. This code also takes into account the serious challenge of foreground contaminants in any highly redshifted $21\,\textrm{cm}$ observation. Foregrounds arise from sources such as Galactic synchrotron radiation, and are four to five orders stronger than the $21\,\textrm{cm}$ cosmological signal in brightness temperature. In this paper, we use the ``moderate foregrounds" setting of {\tt 21cmSense} to account for contamination. This makes the assumption that foregrounds are preferentially confined to certain regions of Fourier space. This confinement is most naturally expressed in terms of spatial Fourier wavenumbers for Fourier modes along the line-of-sight, $k_\parallel$, and wavenumbers for those perpendicular to the line-of-sight, $k_\perp$. Foreground contaminants are expected to appear mostly in modes that satisfy the condition 
\begin{equation}
\label{eq:Wedge}
k_\parallel < k_\parallel^0 + \frac{H_0 D_c \theta_0 \left[\Omega_m (1+z)^3 + \Omega_\Lambda\right]^\frac{1}{2}}{c (1+z)} k_\perp,
\end{equation}
where $c$ is the speed of light, $H_0$ is the Hubble parameter, $D_c$ is the comoving line-of-sight distance, $\Omega_m$ is the normalized matter density, $\Omega_\Lambda$ is the normalized dark energy density, $k_\parallel^0$ is some constant offset, and $\theta_0$ is a characteristic angular scale on the order of the instantaneous field-of-view of radio antennas. Detailed derivations of this formula may be found in, e.g., Ref. \cite{Parsons_et_al2012b,Liu_et_al2014a}, but for the purposes of this paper, it is sufficient to simply understand the qualitative features of this condition, which are as follows. The foregrounds that plague $21\,\textrm{cm}$ experiments are generally expected to possess smooth spectra. Given that redshifted $21\,\textrm{cm}$ observations are mappings of a spectral line, the spectral axis maps to line-of-sight distance $r_\parallel$, and it follows that once they are Fourier transformed, spectrally smooth foregrounds should be seen only at $k_\parallel$ modes below some $k_\parallel^0$ that quantifies the degree of smoothness. However, this is complicated by the inherent chromaticity of interferometers, which may imprint extra spectral structure into the observations of foregrounds, and thus cause them to appear at higher $k_\parallel$. Such effects are particularly pronounced for the longer baselines of an interferometer, which are sensitive to finer spatial structures---higher $k_\perp$ modes---on the sky. This leads to the second term of Eq. \eqref{eq:Wedge}. On the ``moderate foregrounds" setting of the {\tt 21cmSense} code, modes satisfying Eq. \eqref{eq:Wedge} are assumed to be irrecoverably contaminated by foregrounds and are discarded. The power spectrum error bars in the other modes are calculated using the methods of Ref. \cite{parsons_et_al2012a}, where standard formulae for interferometric noise are cast in a cosmological context. At the low frequencies relevant to $21\,\textrm{cm}$ experiments that target reionization, these errors are typically dominated by sky noise, although cosmic variance is also accounted for in {\tt 21cmSense}.

The other category of $21\,\textrm{cm}$ experiments that we consider are known as global signal experiments. Here, the goal is to measure the angle-averaged brightness temperature $\overline{\delta T}_b(\nu)$ as a function of frequency (or equivalently, redshift). As a fiducial experiment, we will consider a single dipole observing the Northern Galactic Pole with a primary beam profile of the form
\begin{equation}
A(\theta, \varphi) = \exp \left( -\frac{1}{2} \frac{\theta^2}{\theta_b^2} \right) \cos \theta,
\end{equation}
where $\theta$ is the polar angle from zenith, $\varphi$ is the azimuthal angle, and $\theta_b$ is a characteristic primary beam width. We take $\theta_b$ to be $0.3\,\textrm{rad}$ at the lowest observation frequency (either $150\,\textrm{MHz}$ or $175\,\textrm{MHz}$ depending on the dataset) and inversely proportional to $\nu$ at other frequencies. Spectral foreground contamination is computed by mock observations of the Global Sky Model of Ref. \cite{deOliveiraCosta_et_al2008}. Observational error bars are computed using the radiometer equation, where the noise temperature variance $\sigma^2$ is given by
\begin{equation}
\sigma^2 = \frac{2 T_\textrm{sys}^2}{t_\textrm{int} \Delta \nu},
\end{equation}
where $t_\textrm{int}$ is the integration time (set to $500\,\textrm{hours}$ for all global signal experiments considered in this paper), $\Delta \nu$ is the frequency channel width (set to $1\,\textrm{MHz}$), and $T_\textrm{sys}$ is the system temperature (set to be equal to the sky temperature for low-frequency, sky-noise dominated regime considered here). The factor of two arises from the squared nature of auto-correlation experiments like the single-dipole experiments considered here, where the variance goes as the four-point function of the (Gaussian-distributed) output voltages.

For the CMB, we make use of publicly available data products from the \emph{Planck} satellite's $2015$ data release. We use only the best fit values for cosmological parameters and their accompanying covariance matrices, essentially approximating parameter uncertainties as being Gaussian, forgoing the also-publicly available non-Gaussian posterior distributions. This approximation is made to match the simplicity of the $21\,\textrm{cm}$ parameter estimates, which are based on the Fisher matrix formalism to avoid the computational expense of a full Bayesian treatment. Throughout the paper, we will focus on the ``TT+lowP" and the ``TT,TE,EE + lowP + lensing + ext" datasets from the \emph{Planck} 2015 data release \cite{Planck2015parameters,Planck2015likelihood}. These datasets bracket the range of uncertainties from the data release, with the TT+low dataset having relatively large errors by \emph{Planck} standards, while TT,TE,EE + lowP + lensing + ext has the tightest error bars. Conveniently, these datasets are also close to representing the extremes in terms of reionization scenarios allowed by CMB data. The TT+low dataset implies a relatively high redshift $z_\textrm{ion}$ for reionization ($z_\textrm{ion} \approx 9.9$, assuming a width $\Delta z_\textrm{ion} \sim 0.5$ in the ionization history), whereas TT,TE,EE + lowP + lensing + ext is best fit by a later reionization epoch ($z_\textrm{ion} \approx 8.8$). As was demonstrated in Ref. \cite{liu_and_parsons2015}, this can have a non-negligible impact on reionization constraints from $21\,\textrm{cm}$ measurements. In either case, HERA's broad frequency range (from $100$ to $200\,\textrm{MHz}$, with strong possibilities for extensions on either end of the spectrum) allows a precise determination of $\tau$ from $21\,\textrm{cm}$ data.

\section{Ingredients for a precise prediction of $\tau$ }
\label{sec:Ingredients}

In practical terms, the optical depth $\tau$ is a nuisance parameter that is self-consistently fit for in CMB studies. While such an approach is attractive in that it does not require detailed models of reionization (or any other process that may produce free electrons), its downside is that one must simply accept any degeneracies in parameter fits. In particular, CMB experiments are much more sensitive to the overall combination of $A_s e^{-2\tau}$ than to $A_s$ or $\tau$ individually. Our goal in this paper is to show how this degeneracy can be broken with the aid of $21\,\textrm{cm}$ data. Typically, this requires modeling the underlying astrophysics of reionization, and in this section we precisely describe the various quantities (both astrophysical and cosmological) that are needed for such modeling.

The optical depth is given by
\begin{equation}
\tau = \sigma_T \int    \overline{n}_e (z)  \frac{dl}{dz} dz,
\end{equation}
where $\sigma_T$ is the Thomson cross-section, $\overline{n}_e$ is the free-electron number density (with the overline denoting an average over all sky directions), and $dl/dz$ is the line-of-sight proper distance per unit redshift. Explicitly, $\overline{n}_e$ may be decomposed as
\begin{eqnarray}
\overline{n}_e &=& \overline{x_\textrm{HII} n_\textrm{H}} + \overline{x_\textrm{HeII} n_\textrm{He}} + \overline{x_\textrm{HeIII} n_\textrm{He}} \nonumber \\
&=& \overline{x_\textrm{HII} n_b} + \frac{1}{4}\overline{x_\textrm{HeIII} n_b}Y_p^\textrm{BBN} \nonumber \\
&=& \overline{n}_b \left[  \overline{x_\textrm{HII} (1+\delta_b)} + \frac{1}{4}\overline{x_\textrm{HeIII} (1+\delta_b)} Y_p^\textrm{BBN} \right],
\end{eqnarray}
where $n_\textrm{H}$, $n_\textrm{He}$, and $n_b = n_\textrm{H} + n_\textrm{He}$ are the hydrogen, helium, and baryon number densities, respectively. The ionization fractions (defined to be between $0$ and $1$) are given by $x_\textrm{HII}$, $x_\textrm{HeII}$, and $x_\textrm{HeIII}$, referring to singly ionized hydrogen, singly ionized helium, and doubly ionized helium, respectively. The helium fraction $Y_p^\textrm{BBN}$ is defined as $4n_\textrm{He} / n_b$, and $\delta_b$ denotes the baryon overdensity.\footnote{We follow the \emph{Planck} team's convention and notation in defining $Y_p^\textrm{BBN}$ as four times the number density fraction, rather than as the helium \emph{mass} fraction (which would instead be defined as $4n_\textrm{He} / [n_\textrm{H} + (m_\textrm{He} / m_\textrm{H}) n_\textrm{H} ]$, where $m_\textrm{H}$ and $m_\textrm{He}$ are the atomic weights of hydrogen and helium, respectively).} In the penultimate equality, we made the standard approximation (justified by simulations \cite{trac_and_cen2007}) that the helium is singly reionized at the same time as hydrogen is, and in the final equality, we used the fact that $n_b =\overline{n}_b ( 1+ \delta_b)$. With this factorization, the averaged baryon density can be easily related to cosmological parameters via
\begin{equation}
\overline{n}_b = \frac{3 H_0^2 \Omega_b}{8 \pi G \mu m_p} (1+z)^3,
\end{equation}
where $\Omega_b$ is the normalized baryon density, $G$ is the gravitational constant, $m_p$ is the mass of the proton, and $\mu$ is the mean molecular weight, which in our case is given by
\begin{equation}
\mu = 1 + \frac{Y_p^\textrm{BBN}}{4} \left( \frac{m_\textrm{He}}{m_\textrm{H}} - 1\right).
\end{equation}
Finally, we assume a flat universe and thus have as our differential line element
\begin{equation}
\frac{dl}{dz} =  \frac{c/H_0}{(1+z) \sqrt{\Omega_m (1+z)^3 + \Omega_\Lambda}}.
\end{equation}

Putting everything together, we may express the total optical depth as $\tau \equiv \tau_\textrm{H} + \tau_\textrm{He}$, with $\tau_\textrm{H}$ and $\tau_\textrm{He}$ denoting the portions of the optical depth sourced by free electrons from HI/HeI reionization and that from HeII reionization, respectively.\footnote{Throughout this paper, we adopt the convention where ``hydrogen reionization" refers to the joint reionization of HI and HeI, whereas ``helium reionization" refers to the ionization of HeII only.} These two contributions take the form
\begin{eqnarray}
\label{eq:tauH}
\tau_\textrm{H} = \frac{3 H_0 \Omega_b \sigma_Tc}{8 \pi G m_p} \left[ 1 + \frac{Y_p^\textrm{BBN}}{4}\left( \frac{m_\textrm{He}}{m_\textrm{H}} - 1\right)\right]^{-1} \nonumber \\
\times \int_0^{z_\textrm{CMB}} \frac{dz (1+z)^2}{\sqrt{\Omega_\Lambda + \Omega_m (1+z)^3}}  \overline{x_\textrm{HII} (1+\delta_b)},
\end{eqnarray} 
and
\begin{eqnarray}
\label{eq:tauHe}
\tau_\textrm{He} = \frac{3 H_0 \Omega_b \sigma_Tc}{8 \pi G m_p} \left[ \frac{4}{Y_p^\textrm{BBN}} + \left( \frac{m_\textrm{He}}{m_\textrm{H}} - 1\right)\right]^{-1} \nonumber \\
\times \int_0^{z_\textrm{CMB}} \frac{dz (1+z)^2}{\sqrt{\Omega_\Lambda + \Omega_m (1+z)^3}}  \overline{x_\textrm{HeIII} (1+\delta_b)},
\end{eqnarray}
where $z_\textrm{CMB}$ is the redshift of the surface of last scattering. From these expressions, we see explicitly how various cosmological parameters and astrophysical fields contribute to a prediction of $\tau$. In what follows, we will discuss the extent to which these contributions must be known accurately before a high-precision value for $\tau$ can be predicted.

\subsection{Uncertainties from fundamental constants and cosmological parameters}
\label{sec:CosmoParamUncertainties}
Eqs. \eqref{eq:tauH} and \eqref{eq:tauHe} both involve a large number of fundamental constants and cosmological parameters, all of which come with their own error bars. Constants such as $G$, $\sigma_T$, $c$, $m_p$, $m_\textrm{H}$, and $m_\textrm{He}$ contribute negligibly to the error budget of $\tau$. The same is true for $Y_p^\textrm{BBN}$, which is constrained to be $0.2467\pm0.0006$ by a combination of \emph{Planck} data and BBN calculations \cite{Planck2015parameters}. The remaining parameters contribute to the error budget in a non-negligible way and must be accounted for.

Consider first the uncertainties arising from cosmological parameters, leaving astrophysical uncertainties in the reionization process to Sec. \ref{sec:astroUncertainties}. To simplify the latter in order to clarify the former, suppose (for this section only) that reionization occurs instantaneously at redshift $z_\textrm{ion}$ (with different values depending on whether one is discussing hydrogen or helium reionization, i.e., whether one is referring to Eqs. \ref{eq:tauH} or \ref{eq:tauHe}). Terms such as $\overline{x_\textrm{HII} (1+\delta_b)}$ and $\overline{x_\textrm{HeIII} (1+\delta_b)}$ thus reduce to step functions that are $1$ for $z<z_\textrm{ion}$ and $0$ otherwise. The integrals in our expressions can then be evaluated analytically, yielding
\begin{equation}
\label{eq:exactConstant}
\tau \propto \frac{h \Omega_b }{\Omega_m} \left[ \sqrt{\Omega_\Lambda + \Omega_m (1+z_\textrm{ion})^3} - 1 \right],
\end{equation}
where we have employed the standard definition $H_0 \equiv 100h \frac{\textrm{km}/\textrm{s}}{\textrm{Mpc}}$, and have omitted the subscripts for $\tau_\textrm{HI}$ and $\tau_\textrm{He}$ in favor of a generic $\tau$ because the dependence on cosmological parameters are the same in either case.

To estimate the uncertainty in this prefactor for $\tau$, we propagate cosmological parameter uncertainties from \emph{Planck} results. To account for error correlations between different parameters, we use the publicly released covariance matrices to draw random samples of $\Omega_b h^2$, $\Omega_c h^2$, and $\theta_\textrm{MC}$, where $\Omega_c$ is the normalized cold dark matter density, and $\theta_\textrm{MC}$ is the {\tt CosmoMC} software package's \cite{lewis_and_bridle2002} approximation to the angular size of the sound horizon at recombination. From this set, all the parameters necessary for evaluating Eq. \eqref{eq:exactConstant} can be obtained. Using \emph{Planck's} TT + lowP covariance from the 2015 data release (featuring relatively high $\Omega_m$ and $\tau$), the fractional error in Eq. \eqref{eq:exactConstant} is $1.40\%$. Similar results are obtained for TT,TE,EE + lowP + lensing + ext (featuring relatively low $\Omega_m$ and $\tau$), with a fractional error of $0.75\%$. Note that these values are merely rough estimates of how cosmological parameter uncertainties can affect a $21\,\textrm{cm}$-derived prediction of $\tau$. This is because we have so far only considered the influence that cosmological parameters have on the ``geometric" portions of $\tau$ (e.g., $dl/dz$). In reality, cosmological parameters also affect quantities such as $x_\textrm{HII}$, leading to the possibility that the final errors may be different from what is predicted in this section. However, it is reassuring that our estimates are small enough that it appears to be a worthwhile exercise to use $21\,\textrm{cm}$ observations to better constrain $\tau$. We will find this conclusion to be unchanged when we include the non-geometric influence of cosmological parameters in Sec. \ref{eq:Pkformalism}.

\subsection{Uncertainties from astrophysical processes}
\label{sec:astroUncertainties}

We now consider the uncertainties in predicting $\tau$ that arise directly from uncertain astrophysics (aside from the subtler changes to astrophysics occurring because of shifts in cosmological parameters that we alluded to above). At the crudest level, changes in astrophysics affect the redshift of reionization $z_\textrm{ion}$, which affects the optical depth via Eq. \eqref{eq:exactConstant}. Indeed, working in reverse and solving for $z_\textrm{ion}$ given a measured value of $\tau$ is how CMB experiments have traditionally placed constraints on reionization, although recent advances in higher order effects such as the kinetic Sunyaev-Zel'dovich effect have enabled increasingly sophisticated limits \cite{zahn_et_al2012,sievers_et_al2013,george_et_al2015}.

\begin{figure}[!]
	\centering
	\includegraphics[width=0.5\textwidth,trim=0.5cm 0cm 0.75cm 0.90cm,clip]{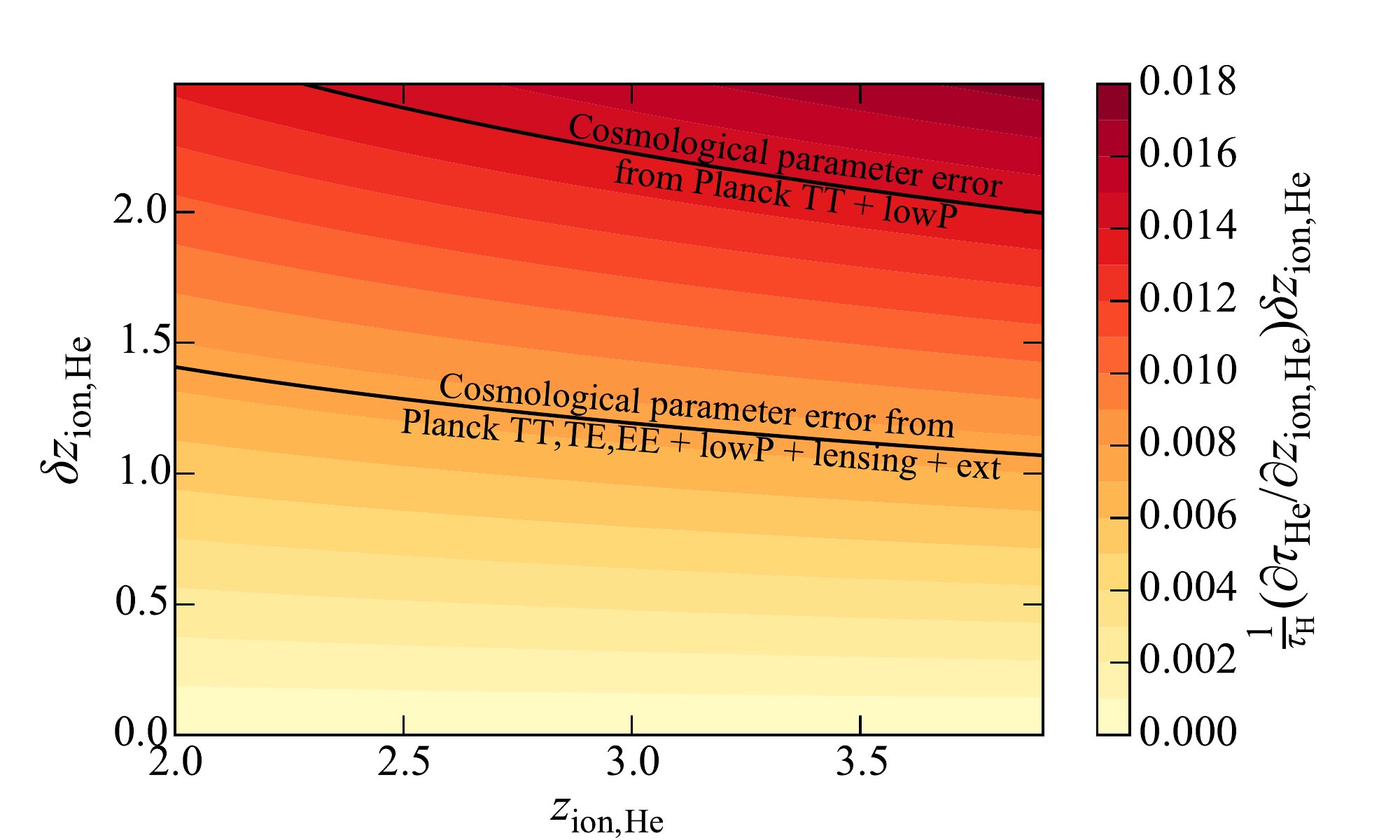}
	\caption{Fractional error in $\tau$ induced by uncertainties in helium reionization as a function of the redshift of helium reionization $z_\textrm{ion,He}$ and the uncertainty in this redshift $\delta z_\textrm{ion,He}$. For reasonable values of these parameters, the errors arising from uncertainties in helium reionization are subdominant to those arising from cosmological parameter uncertainty. It is thus permissible to neglect uncertainties in helium reionization.}
	\label{fig:HeII_reion_errors}
\end{figure}

Ultimately, we shall see that it is important to model reionization astrophysics in detail, beyond the simple parametrization of $z_\textrm{ion}$. However, by considering the coarse dependence of $z_\textrm{ion}$ on $\tau$, we can distinguish the pieces of astrophysics that need to be carefully modeled from those that do not. In particular, we will now show that helium reionization contributes relatively little to $\tau$, making simple models of the process sufficient. Consider the ratio of $\tau_\textrm{He}$ to $\tau_\textrm{H}$, which can be written as
\begin{equation}
\frac{\tau_\textrm{He}}{\tau_\textrm{H}} = \frac{Y_p^\textrm{BBN}}{4} \left[ \frac{ \sqrt{\Omega_\Lambda + \Omega_m (1+z_\textrm{ion,He})^3} - 1}{ \sqrt{\Omega_\Lambda + \Omega_m (1+z_\textrm{ion,H})^3} - 1}\right],
\end{equation}
where $z_\textrm{ion,He}$ and $z_\textrm{ion,H}$ are the redshifts of helium and hydrogen reionization, respectively, assuming that both processes are instantaneous. For a fiducial model with $\Omega_\Lambda = 0.6911$, $\Omega_m = 0.3089$, $z_\textrm{ion,H} = 8.8$ (corresponding to \emph{Planck's} TT,TE,EE + lowP + lensing + ext dataset), and $z_\textrm{ion,He} = 3$, this ratio is $\sim 1.4\%$. Any errors in helium reionization are then suppressed by this factor. Quantitatively, if we parameterize the uncertainty in helium reionization by considering a shift $\delta z_\textrm{ion,He}$ in $z_\textrm{ion,He}$, the fractional error in $\tau$ arising from such uncertainty is given approximately by $\tau_\textrm{ion,H}^{-1} (\partial \tau_\textrm{He} / \partial z_\textrm{ion,He}) \delta z_\textrm{ion,He}$. This quantity is shown in Fig. \ref{fig:HeII_reion_errors}, where we have overlaid the fractional errors from cosmological parameter uncertainties that we computed in the previous section. One sees that so long as the redshift of helium reionization $z_\textrm{ion,He}$ can be constrained reasonably well (say, $ \delta z_\textrm{ion,He} \sim 1$), the uncertainties in the astrophysics of helium reionization are subdominant to those from cosmological parameters. As a proxy for $ \delta z_\textrm{ion,He} $, consider the spread in the inferred $z_\textrm{ion,He}$ values from HeII Ly$\alpha$ absorption studies. With an increased number of sightlines, recent studies have shown significant scatter in $z_\textrm{ion,He}$, suggesting a rather extended helium reionization between $z \sim 3.4$ and $z \sim 2.7$ \cite{worseck_et_al2014}, but this still satisfies our requirement that $ \delta z_\textrm{ion,He} \lesssim 1$. Admittedly, this is a rather crude way to estimate error contributions from helium reionization, one which can be easily improved upon even with current data. For now, however, we will assume for simplicity that uncertainties in helium reionization can be ignored based on our back-of-the-envelope estimates.

\begin{figure}[!]
	\centering
	\includegraphics[width=0.45\textwidth]{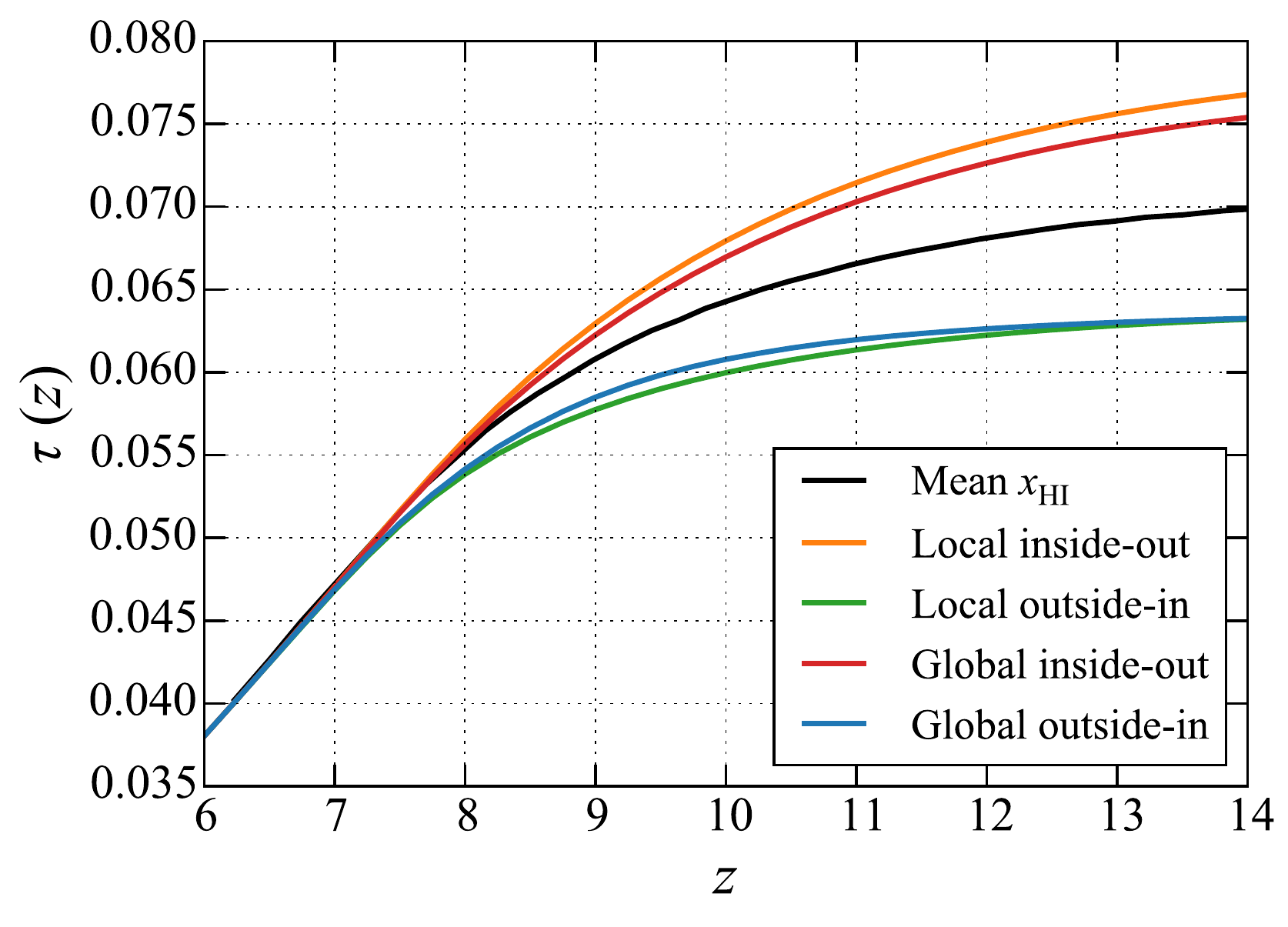}
	\caption{Cumulative contribution to the optical depth $\tau$ from low to high redshift, for several different models of reionization. The crucial astrophysical quantity for a precise determination of $\tau$ is the density-weighted ionized fraction. This depends on the correlation between the ionization field and the density field. The different reionization models shown here reflect different models for this correlation, which must be known for a precise prediction of $\tau$, given the spread seen here.}
	\label{fig:InsideOutvsOutsideIn}
\end{figure}

In contrast, the astrophysics of hydrogen reionization must be accurately modeled for precise predictions of $\tau$. Repeating the above analysis for order unity perturbations in the $z_\textrm{ion,H}$, the resulting change in $\tau$ is $\sim 17\%$, largely because there is no longer a suppression by the ratio $\tau_\textrm{He} / \tau_\textrm{H}$. Of course, this is hardly surprising, for if changes in $z_\textrm{ion,H}$ did not generate reasonably large shifts in $\tau$, CMB-derived constraints on reionization would not exist. For our goal of predicting $\tau$ to be worthwhile, then, the details of hydrogen reionization must be understood. In fact, with hydrogen reionization dominating the CMB optical depth, one must also go beyond simple models of instantaneous reionization. To see this, consider the following numerical experiment. The astrophysics of reionization enters Eq. \eqref{eq:tauH} via the $\overline{x_\textrm{HII} (1+\delta_b)}$ term, the density-weighted ionized fraction. Crucially, it is incorrect to simplify this term to $\overline{x}_\textrm{HII}\overline{ (1+\delta_b)}$ (which would consequently make it equal to $\overline{x}_\textrm{HII}$), since $x_\textrm{HII}$ and $\delta_b$ may be spatially correlated, making the angular average of their product different from the product of their averages. In general, spatial correlations are an expected feature of reionization. For example, in ``inside-out" models of reionization, higher density regions produce a greater number of ionized photons and preferentially ionize first \cite{barkana_and_loeb2004,furlanetto_et_al2006}, resulting in a positive correlation between $x_\textrm{HII}$ and $\delta_b$. This is in contrast to ``outside-in" models, where recombinations limit the rate of ionization, and thus higher density regions (where recombinations are more common) are ionized last \cite{Miralda-Escude_et_al2000}. This results in a negative correlation between ionization and density. Fig. \ref{fig:InsideOutvsOutsideIn} shows the differential contributions to the total optical depth in various models, all with the same mean ionization history $\overline{x}_\textrm{HII} (z) $. These are based on simulations used in Ref. \cite{watkinson_and_pritchard2014}, where the interested reader will be able to find details. In brief, the curves labeled ``Global" refer to reionization models where the morphology of reionization is driven by the large scale structure of the density field. This tends to lead to large ionized regions around clustered high density peaks. On the other hand, the curves labeled ``Local" refer to reionization models where small ionized bubbles form around individual galaxies. Each type of reionization morphology (``Global" or ``Local") is then also paired with the inside-out or outside-in scenarios discussed above. One sees from Fig. \ref{fig:InsideOutvsOutsideIn} that these details of reionization matter, and induce errors of $\sim 5$ to $10\%$ in $\tau$ if not accounted for.

Mathematically, the fact that $\overline{x_\textrm{HII} (1+\delta_b) }$ does not reduce to $\overline{x}_\textrm{HII}$ is due to our use of angular averages, which are akin to volume averages. This makes our definition of the ionized fraction $\overline{x}_\textrm{HII}$ conform to the convention typically employed in the reionization literature, where it is often denoted the volume filling factor of ionized fraction. Such a definition is convenient for relating our measurements to simulations (as we do so in the following section), since the simulations provide ionization fractions in cells of fixed volume. If one prefers, it is of course permissible to define one's averages in terms of mass-averages. Eq. \eqref{eq:tauH} then amounts to an integral (with appropriate geometric factors) over $\overline{x}_\textrm{HII}$. In some sense, though, this is merely a cosmetic change, for in order to accurately compute a mass-weighted average, the reionization simulations must still model the spatial correlations between the density and ionization.

In summary, the uncertainties in $\tau$ predictions arise from both uncertainties in cosmological parameters and uncertainties in astrophysics. If not accounted for in detail, the astrophysics contributes more to errors in $\tau$ than the cosmology does. To make progress, then, our goal is to use $21\,\textrm{cm}$ data to better understand the astrophysics of reionization.

\section{Relating $\tau$ to HI surveys}
\label{eq:Pkformalism}

As we have seen in previous sections, predictions of $\tau$ are currently dominated by uncertainties in astrophysics. In this section, we establish formalism for incorporating $21\,\textrm{cm}$-derived astrophysical constraints from reionization to provide better measurements of $\tau$ (and thus other cosmological parameters) than one can obtain using the CMB alone.

The brightness temperature contrast $\delta T_b$ of the redshifted $21\,\textrm{cm}$ line against the CMB is given by \cite{furlanetto_et_al2006,aviBook}
\begin{equation}
\label{eq:deltaTdef}
\delta T_b(\mathbf{\hat{n}}, \nu) \approx \delta T_{b0}\, x_\textrm{HI}  (1 + \delta_b) \left( 1 - \frac{T_\gamma}{T_s}\right)\!\left( \frac{H}{H+ \partial v_r / \partial r} \right),
\end{equation}
with
\begin{eqnarray}
\label{eq:TbPrefactor}
\delta T_{b0} &=&  \frac{9\hbar c^2 A_{10}\Omega_b H_0}{128 \pi G k_B \nu_{21}^2 \mu m_p\Omega_m^{1/2}} \left(1 - \frac{Y_p^\textrm{BBN}}{4} \right) \nonumber \\
&\approx& 28 \left( \frac{1+z}{10} \frac{0.14}{\Omega_m h^2} \right)^\frac{1}{2} \left( \frac{\Omega_b h^2}{0.022}\right)\,\textrm{mK},
\end{eqnarray}
where $\hbar$ is the reduced Planck's constant, $A_{10}=2.85\times10^{-15}\,\textrm{s}^{-1}$ is the spontaneous emission coefficient of the $21\,\textrm{cm}$ transition, $k_B$ is Boltzmann's constant, $\nu_\textrm{21} \approx 1420\,\textrm{MHz}$ is the frequency of the $21\,\textrm{cm}$ line, $x_\textrm{HI} = 1 - x_\textrm{HII}$ is the hydrogen neutral fraction, $T_\gamma$ is the temperature of the CMB, $T_s$ is the spin temperature of the hydrogen atoms, and $\partial v_r / \partial r$ is the derivative of the comoving radial peculiar velocity with respect to the comoving radial distance. The peculiar velocity gradient is assumed to be small relative to the Hubble parameter $H$ in Eq. \eqref{eq:TbPrefactor}, and it is understood that $T_\gamma$, $T_s$, $\delta_b$, $x_\textrm{HI}$, $H$, and $\partial v_r / \partial r$ are evaluated at redshift $z = (\nu_\textrm{21} / \nu ) -1$. The brightness temperature field is sensitive to both the cosmology (via $\delta_b$, $H$, $\partial v_r / \partial r$ and standard cosmological parameters) and the astrophysics (via $x_\textrm{HI}$ and $T_s$) of reionization. Since the product of $x_\textrm{HI} \delta_b$ enters the expression for $\delta T_b$, the $21\,\textrm{cm}$ line is clearly sensitive to the correlations between density and ionization, which we argued in the previous section are a crucial ingredient in our quest to understand reionization well enough to precisely predict $\tau$.

%

To harness the $21\,\textrm{cm}$ line for a $\tau$ prediction, however, there are two challenges that must be overcome. First, it is necessary to make redshifted $21\,\textrm{cm}$ measurements that have high enough signal-to-noise to be useful. Unfortunately, a combination of sensitivity limitations and foreground contamination make direct mapping of the brightness temperature field unlikely in the near future. More observationally attainable in the short term are measurements of the brightness temperature power spectrum $P_\textrm{21} (k)$, as defined by Eq. \eqref{eq:PspecDef}.

\begin{figure*}[!]
	\centering
	\includegraphics[width=1.0\textwidth]{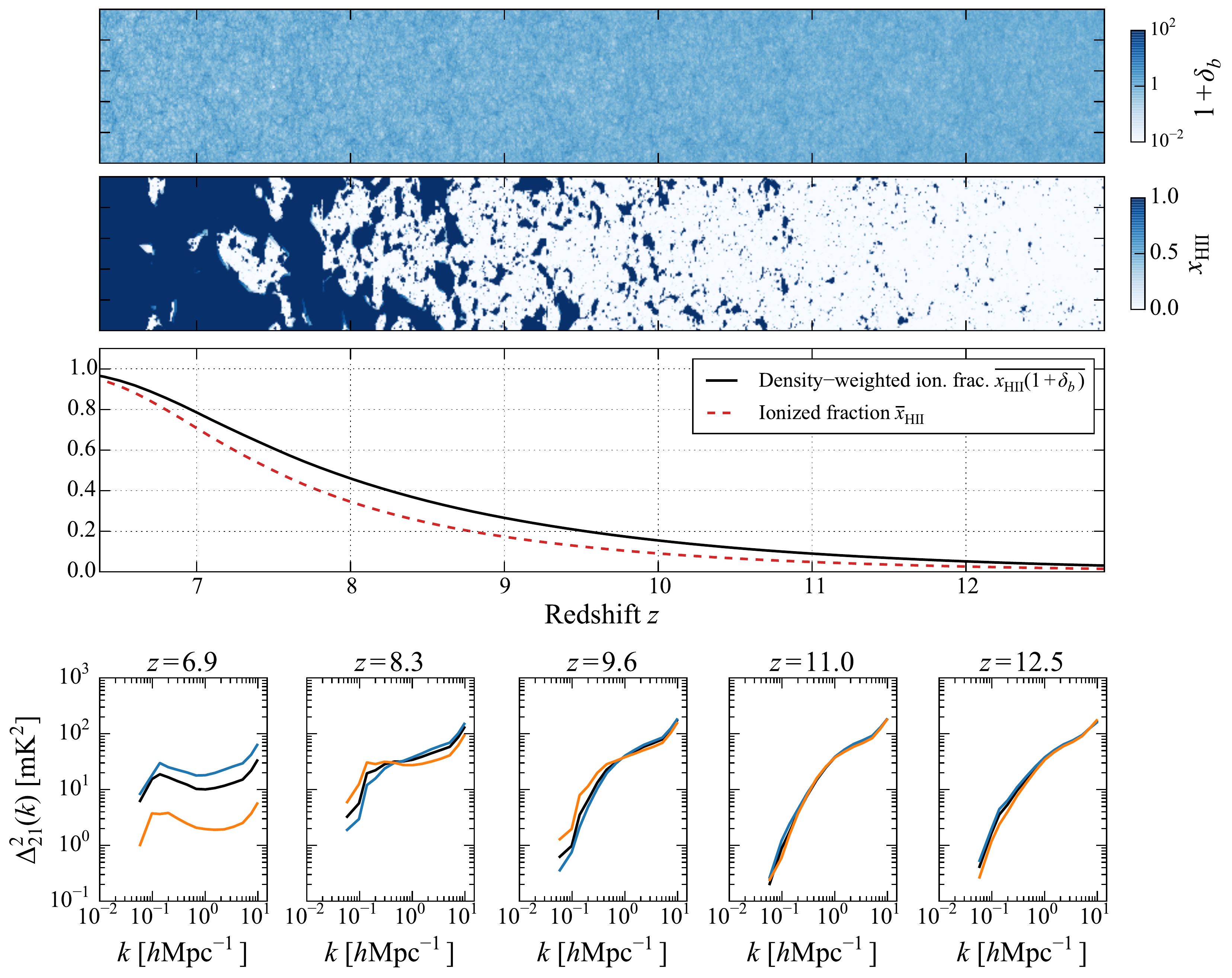}
	\caption{Top row: Simulation of the nonlinear density field over the past light cone that is observed by a $21\,\textrm{cm}$ experiment. Second row: Corresponding ionization fraction, assuming $(T_\textrm{vir}, R_\textrm{mfp}, \zeta) = (6 \times 10^4\,\textrm{K}, 35\,\textrm{Mpc}, 30)$ to match the optical depth of \emph{Planck} TT,TE,EE + lowP + lensing + ext. Third row: Corresponding ionized fraction history $\overline{x}_\textrm{HII}$ (red solid curve) and the density-weighted ionization history $\overline{x_\textrm{HII} (1+ \delta_b)}$ (black solid curve). The averaged ionized fraction is also seen to be a poor approximation for the density-weighted ionized fraction, which is the crucial quantity for determining $\tau$. Bottom row: Corresponding $21\,\textrm{cm}$ power spectra (black) at various redshifts, plotted as $\Delta^2_{21} (k) \equiv k^3 P_{21}(k) / 2 \pi^2$. Blue and orange curves show power spectra for different values of $T_\textrm{vir}$. Note that this figure is intended for illustrative purposes only, and that the scales on the top two rows do not correspond exactly to the redshift axis on the third row. In our proposed analysis, one measures the bottom row through observations, constraining underlying model parameters that are then fed into simulations to produce the top two rows. The density-weighted ionization fraction (third row) is then extracted and inserted into Eq. \eqref{eq:tauH} to determine $\tau$.}
	\label{fig:Simulations}
\end{figure*}

Having identified the $21\,\textrm{cm}$ power spectrum as a promising near-term, high signal-to-noise measurement of reionization, the second challenge is the translation of our measurements into a precise prediction of $\tau$. Fundamentally, what is needed for Eq. \eqref{eq:tauH} is the density-weighted ionized fraction, but as one sees from substituting Eq. \eqref{eq:deltaTdef} into Eq. \eqref{eq:PspecDef}, the $21\,\textrm{cm}$ power spectrum probes a much more complicated combination of parameters and their correlations. To connect our power spectrum observations to the underlying fields needed for our $\tau$ prediction, we appeal to semi-analytic simulations. In particular, we assume an inside-out model of reionization based on the excursion set formalism of Ref. \cite{furlanetto_et_al2004}, as implemented in the publicly available {\tt 21cmFAST} software \cite{mesinger_et_al2011}. Of course, this is a rather specific model of reionization, and once $21\,\textrm{cm}$ measurements move beyond an initial detection, it will be crucial to test the validity of the model. For simplicity, we will assume for the rest of the paper that such a model selection exercise has already been performed, which upcoming instruments should be able to do at high significance \cite{watkinson_and_pritchard2014}. We base our forecasts on the excursion set-based inside-out model of {\tt 21cmFAST} because it runs quickly and has been shown to agree reasonably well with state-of-the-art ray-tracing radiative transfer simulations \cite{zahn_et_al2011}, although this comes with the caveat that the agreement is best when the comparison is made at equal ionization fractions, not equal redshifts. For computational convenience we run {\tt 21cmFAST} in the mode where $T_s \gg T_\gamma$. This is expected to be a good approximation after the very beginning of reionization \cite{santos_et_al2008}, making it a suitable simplifying assumption for us to employ given the redshift ranges that we assume for power spectrum observations (detailed below). However, we stress that our formalism can be applied to any reionization simulation; in what follows, one simply replaces Eqs. \eqref{eq:TTlowP_linearTau} and \eqref{eq:TTTEEE_linearTau} with expressions calibrated to a chosen simulation.

Fig. \ref{fig:Simulations} illustrates how semi-analytic simulations can be used to connect $21\,\textrm{cm}$ power spectrum measurements to $\tau$. The bottom row of the figure shows the $21\,\textrm{cm}$ power spectra at various redshifts, plotted as $\Delta^2_{21} (k) \equiv k^3 P_{21}(k) / 2 \pi^2$. These are (after some data analysis) what $21\,\textrm{cm}$ experiments measure.\footnote{For simplicity, we leave light cone effects in $21\,\textrm{cm}$ measurements to future work, although such effects should ultimately be taken into account \cite{barkana_and_loeb2006,datta_et_al2012,datta_et_al2014,laplante_et_al2014,zawada_et_al2014,ghara_et_al2015}} With power spectra in hand, one can simultaneously fit for astrophysical and cosmological parameters in an underlying model of reionization, using priors from other cosmological probes such as the CMB. Once the underlying parameters have been determined, semi-analytic simulations can be run to produce past light cone maps of the non-linear baryon density (top row of figure) and ionization fields and ionization fields (second row). These maps can then be used to form $\overline{x_\textrm{HII} (1+ \delta_b)}$ (third row), which is then inserted into Eq. \eqref{eq:tauH} to predict $\tau$. Note from Fig. \ref{fig:Simulations} that there is a clear difference between the $\overline{x}_\textrm{HI}$ and $\overline{x_\textrm{HII} (1+ \delta_b)}$ curves, again illustrating the importance of modeling correlations between density and ionization.

Importantly, the semi-analytic simulations used for the procedure outlined above must span a wide range of redshifts, from before reionization has begun to after reionization is complete. This is necessary because even small levels of ionization can perturb the predicted value of $\tau$ by more than our final error bars. By producing full histories of the density and ionization fields, the simulations compensate for the limited reach in redshift of near-term experiments, which are unlikely to probe the very beginning of reionization to high precision. The self-consistency required by a simulation produces a full reionization history once the model parameters are fixed by (relatively) low-redshift observations. This extrapolation does come with some uncertainty, which we address in Sec. \ref{sec:ModelDependence}.

\begin{figure}[!]
	\centering
	\includegraphics[width=0.5\textwidth]{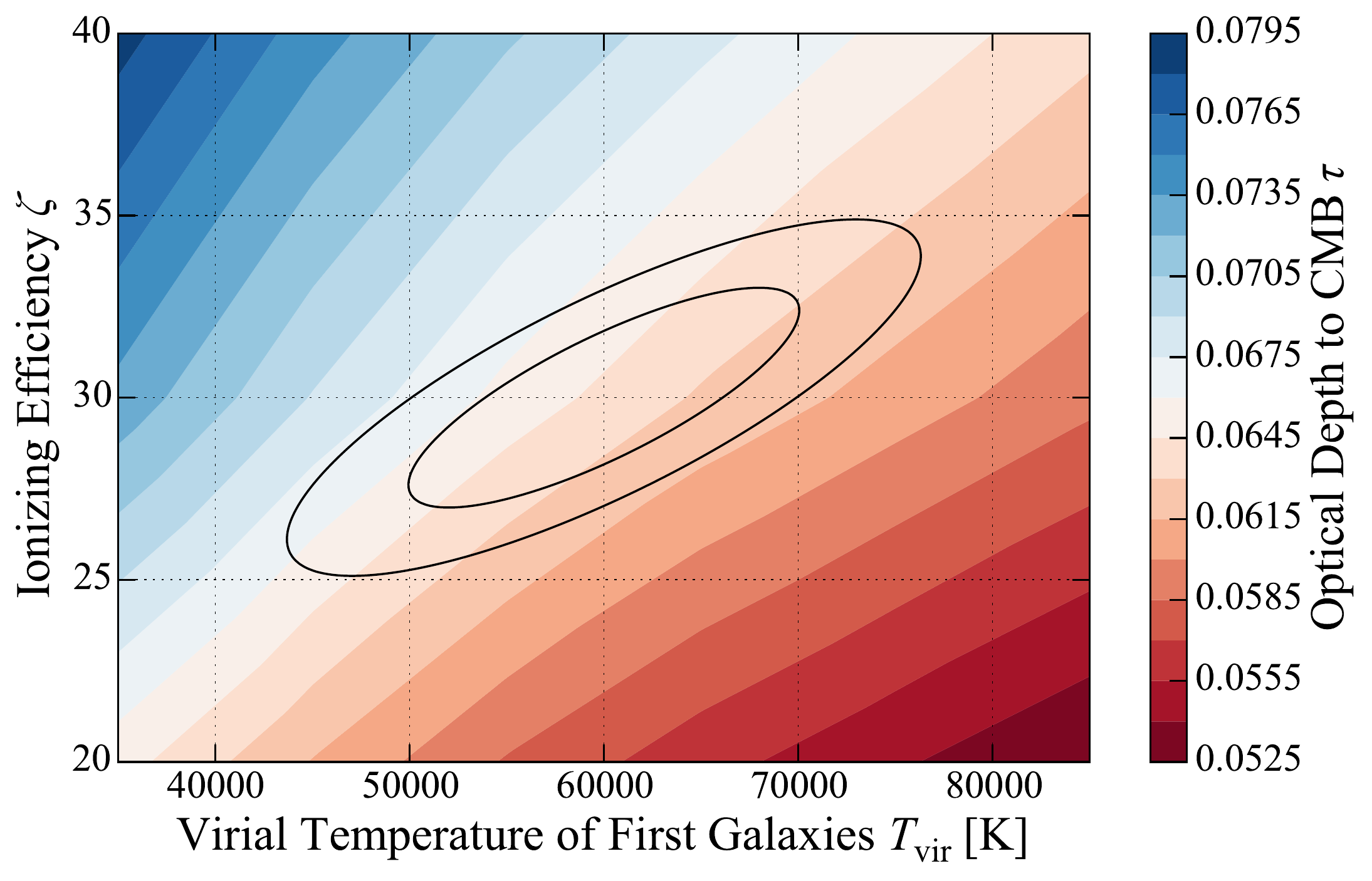}
	\caption{Forecasted $68\%$ and $95\%$ confidence regions (black ellipses) in the $T_\textrm{vir}$-$\zeta$ parameter space for HERA observations, along with {\tt 21cmFAST}-predicted optical depth $\tau$ (filled color contours). The rough alignment of the degeneracy directions suggest that uncertainties in astrophysical parameters arising from $21\,\textrm{cm}$ power spectrum measurements are unlikely to seriously compromise one's ability to make highly precise predictions of $\tau$.}
	\label{fig:21cmDegen_wTau}
\end{figure}

Another limitation of our observations lies in the inherent uncertainty of a $21\,\textrm{cm}$ power spectrum measurement. Instrumental noise, foregrounds, and degeneracies inherent in analyzing a power spectrum observation mean that our predicted value of $\tau$ will come with a corresponding set of uncertainties. Fortunately, we will now see that these errors and degeneracies in $21\,\textrm{cm}$ measurements are unlikely to seriously compromise out ability to predict $\tau$. Following Refs. \cite{mesinger_et_al2012,pober_et_al2014}, we consider a three-parameter model of reionization, parameterized by $T_\textrm{vir}$, the minimum virial temperature of the first ionizing galaxies; $\zeta$, the ionizing efficiency of those galaxies; and $R_\textrm{mfp}$, the mean free path of ionizing photons in ionized regions of our Universe.\footnote{Of course, there exist a large number of other parameterizations and models to describe reionization (e.g., \cite{zahn_et_al2011,battaglia_et_al2013,laplante_et_al2014,gnedin_et_al2014,kaurov_et_al2015a,kaurov2015}). Our intention here is not to imply that the three-parameter model employed here will be applicable to future $21\,\textrm{cm}$ measurements without modification. Instead, it is simply a model that is designed to be both reasonably realistic and flexible enough to encompass a large variety of reionization histories \cite{mesinger_et_al2012}.} As shown in Ref. \cite{pober_et_al2014}, $21\,\textrm{cm}$ power spectrum measurements tend to constrain these parameters in a way that leaves $T_\textrm{vir}$ and $\zeta$ largely degenerate. While multi-redshift information does help to break this degeneracy, it tends to remain to some degree. This can be seen in Fig. \ref{fig:21cmDegen_wTau}, where we show Fisher matrix projections for parameter constraints on $T_\textrm{vir}$ and $\zeta$. The black contours demarcate the $68\%$ and $95\%$ confidence regions for a hypothetical power spectrum measurement performed by HERA spanning $6.1 \leq z \leq 9.1$ at intervals of $\Delta z = 0.3$ (chosen roughly to give good parameter constraints \cite{liu_and_parsons2015}). These are calculated by first computing power spectrum sensitivities using {\tt 21cmSense}, which are then fed into a Fisher matrix computation based on that employed in Ref. \cite{pober_et_al2014}, except with our current experimental parameters and fiducial astrophysical parameters set at $(T_\textrm{vir}, R_\textrm{mfp}, \zeta) = (6 \times 10^4\,\textrm{K}, 35\,\textrm{Mpc}, 30)$. Our value of $T_\textrm{vir}$ corresponds to a virial mass of order $\sim 10^9 M_\odot$ at $z=9$, consistent with typical values adopted elsewhere in the literature. The fiducial values for our three astrophysical parameters are chosen to match the best-fit optical depth of $\tau =0.066$ from the \emph{Planck} TT,TE,EE + lowP + lensing + ext dataset. Marginalizing over cosmological parameters as well as $R_\textrm{mfp}$ gives the contours in Fig. \ref{fig:21cmDegen_wTau}.

For every point in $T_\textrm{vir}$-$\zeta$ space we also show values for $\tau$, predicted from {\tt 21cmFAST} using the procedure outlined above with cosmological parameters fixed at their fiducial values. Here and in the rest of the paper, we assume that helium is instantaneously reionized at $z = 3$, having argued earlier that uncertainties in helium reionization are negligible. Immediately striking is the way in which contours of constant $\tau$ are roughly aligned with the contours from our power spectrum constraint on $T_\textrm{vir}$ and $\zeta$. This would be bad news if our goal was to use CMB measurements of $\tau$ to place additional constraints on the astrophysical parameters of reionization, since parallel contours mean that the constraints are not complementary, not to mention the fact that the $68\%$ confidence interval on $\tau$ from the CMB roughly spans the entire color scale of Fig. \ref{fig:21cmDegen_wTau}. However, parallel contours are desirable for the goals of this paper, since they mean that the inherent degeneracies in one's ability to predict reionization parameters from $21\,\textrm{cm}$ power spectrum measurements do not detract from one's ability to make a highly precise prediction of $\tau$.

Fundamentally, this rather fortunate alignment of contours arises because both the $21\,\textrm{cm}$ line and $\tau$ are probes of reionization that are particularly sensitive to the \emph{timing} of the process, but are relatively insensitive to parameter shifts that leave the timing the same. Consider a simultaneous increase in $T_\textrm{vir}$ and $\zeta$, for example. Increasing $T_\textrm{vir}$ means that reionization is driven by more massive galaxies, which are fewer in number. If one correspondingly increases $\zeta$, however, each galaxy within this rarer population will produce more ionizing photons, leaving the timing of reionization roughly unchanged. The result will have little impact on $\tau$, which is only affected by the total column density of free electrons between us and the surface of last scattering, with no regard for whether these free electrons were produced by a large population of faint ionizing sources or a small population of bright sources. As for the power spectrum measurements, previous work \cite{pober_et_al2014} has shown that redshift evolution is one of the principal ways to break astrophysical parameter degeneracies. Combinations of parameter shifts that leave timing unchanged therefore survive as the residual degeneracies seen in Fig. \ref{fig:21cmDegen_wTau}.

Importantly, the similarities between constraints from $\tau$ and those from $21\,\textrm{cm}$  power spectra seem to be generally robust. Switching to \emph{Planck's} TT+lowP dataset, we match its best-fit $\tau$ value of $0.078$ by using a fiducial astrophysical parameter set of $(T_\textrm{vir}, R_\textrm{mfp}, \zeta) = (4 \times 10^4\,\textrm{K}, 35\,\textrm{Mpc}, 40)$. To better match the higher redshift of reionization, we also adjust our experimental parameters by assuming that observations are analyzed in $\Delta z = 0.5$ portions spanning the range $7.5 \le z \le 10.5$ \cite{liu_and_parsons2015}. We find the same qualitative effects as we did for \emph{Planck's} TT,TE,EE + lowP + lensing + ext dataset.

Proceeding with our prediction of $\tau$, it is crucial to incorporate cosmological parameter uncertainties into our estimate (particular those from $\Omega_m$ and $\Omega_b$). As we saw in Sec. \ref{sec:CosmoParamUncertainties}, cosmological parameter uncertainties can induce roughly percent level errors in $\tau$, which will turn out to be a substantial fraction of the error budget in our final predictions. It is thus incorrect to simply integrate over the likelihood contours Fig. \ref{fig:21cmDegen_wTau} against the values of $\tau$, for those values were computed assuming fixed cosmological parameters. It is also essential to go beyond the approach of Sec. \ref{sec:CosmoParamUncertainties}, where our assumption of instantaneous reionization meant that cosmological parameters only entered ``geometrically" via the prefactors of Eqs. \eqref{eq:tauH} and \eqref{eq:tauHe}. In our more detailed treatment here, we expect $\overline{x_\textrm{HII} (1+\delta_b)}$ to depend on both cosmological and astrophysical parameters.

Suppose we define a function $\tau_\textrm{sim} (\mathbf{p})$ that returns the value of $\tau$ from our simulations given a set of parameters $\mathbf{p}$. For most of this paper (our later discussion of the neutrino mass being an exception), we will pick the three reionization parameters described above, plus the base $\Lambda$CDM parameters used by \emph{Planck} but without $\tau$, i.e., $\mathbf{p}~=~\left[ \Omega_b h^2, \Omega_c h^2, 100\theta_\textrm{MC}, \ln ( 10^{10} A_s), n_s, T_\textrm{vir}, R_\textrm{mfp}, \zeta \right]$, where $\Omega_c$ is the normalized cold dark matter density, $A_s$ is the amplitude of the primordial curvature power spectrum, $n_s$ is the scalar spectral index, and the other parameters retain their definitions from earlier in the paper. In general, $\tau_\textrm{sim}$ is a complicated function of $\mathbf{p}$, and it is computationally impractical to evaluate it directly in (for example) a likelihood analysis.\footnote{Recent efforts in Ref. \cite{greig_and_mesinger2015} have shown that full Bayesian analyses are viable if only astrophysical parameters are varied. However, incorporating cosmological parameter variations into such analyses will require further speed-ups of semi-analytic simulations.}  In practice, however, it is sufficient to simply linearize the relation, since we need not understand how $\tau_\textrm{sim}$ varies over all possible parameter values. Instead, it is only necessary to consider variations induced by perturbations within the narrow ranges of cosmological parameters allowed by \emph{Planck} and astrophysical parameters in soon-to-exist $21\,\textrm{cm}$ results. Degeneracies in the $21\,\textrm{cm}$ results (such as the $T_\textrm{vir}$-$\zeta$ degeneracy discussed above) are of little concern since we have shown that such degeneracies have little effect on $\tau$. If we denote by $\Delta p$ the perturbation of parameter $p$ about its fiducial value in our simulations (not to be confused with the error bar for $p$ in measurements), we find that numerical fits to $\tau_\textrm{sim}$ yield
\begin{eqnarray}
\label{eq:TTlowP_linearTau}
\tau_\textrm{sim}^\textrm{TT+lowP} \approx &&\, 0.078 + 0.042 \left(\frac{\Delta\Omega_b h^2}{0.02222}\right) +  0.11\left(\frac{\Delta\Omega_c h^2}{0.1197}\right) \nonumber\\
&& -0.0074 \left(\frac{\Delta100\theta_\textrm{MC}}{1.04085}\right)  + 0.22  \left(\frac{\Delta\ln (10^{10} A_s)}{3.089}\right)\nonumber \\
&& + 0.27 \left(\frac{\Delta n_s}{0.9655}\right) -0.018\left(\frac{\Delta T_\textrm{vir}}{4 \times 10^4\,\textrm{K}}\right)\nonumber \\
&& -0.0011 \left(\frac{\Delta R_\textrm{mfp}}{35\,\textrm{Mpc}}\right) + 0.020 \left(\frac{\Delta \zeta}{40\,\textrm{Mpc}}\right)
\end{eqnarray}
for the fiducial model based on \emph{Planck's} TT+lowP dataset, and
\begin{eqnarray}
\label{eq:TTTEEE_linearTau}
\tau_\textrm{sim}^\textrm{TT,TE,\dots} \approx &&\, 0.064 + 0.033  \left(\frac{\Delta\Omega_b h^2}{0.02230}\right) + 0.088\left(\frac{\Delta\Omega_c h^2}{0.1188}\right) \nonumber\\ 
&&  -0.0075 \left(\frac{\Delta100\theta_\textrm{MC}}{1.04093}\right) +   0.18\left(\frac{\Delta\ln (10^{10} A_s)}{3.064}\right)\nonumber \\
&& +   0.21 \left(\frac{\Delta n_s}{0.9667}\right)  -0.017 \left(\frac{\Delta T_\textrm{vir}}{6 \times 10^4\,\textrm{K}}\right)\nonumber \\
&& -0.00099 \left(\frac{\Delta R_\textrm{mfp}}{35\,\textrm{Mpc}}\right) +  0.018 \left(\frac{\Delta \zeta}{30\,\textrm{Mpc}}\right)
\end{eqnarray}
for the fiducial model based on \emph{Planck's} TT,TE,EE + lowP + lensing + ext dataset. Drawing $50$ random samples from the final likelihood function (derived at the end of this section) obtained from combining $21\,\textrm{cm}$ data with CMB data, we find that the maximum error in our linear approximation for $\tau$ to be $0.6\%$, and the mean error to be $0.2\%$. Since we have (arbitrarily) scaled each perturbation to the fiducial parameter values, the coefficients of each term in these relations can be interpreted as the change induced in $\tau$ per fractional shift in parameter values. Examining the relative magnitudes of these coefficients, one sees yet more evidence that cosmological parameters can have a significant effect on a $\tau$ prediction.

Because tensions have often arisen in the best fit values of the Hubble parameter $h$ from different datasets, it is worthwhile to consider the dependence of $\tau_\textrm{sim}$ on $h$. Eliminating $\theta_\textrm{MC}$ in favor of $h$ in our linearized relations, one obtains
\begin{equation}
\tau_\textrm{sim}^\textrm{TT+lowP} \approx 0.078  -0.0015 \left(\frac{\Delta h}{0.6731}\right) + \dots 
\end{equation}
and
\begin{equation}
\tau_\textrm{sim}^\textrm{TT,TE,\dots} \approx 0.064  -0.0015 \left(\frac{\Delta h}{0.6774}\right) + \dots ,
\end{equation}
where we have omitted other terms because their coefficients change by very small amounts (as is the case with $\Omega_b h^2$ and $\Omega_c h^2$) or not at all because they are unrelated to $\theta_\textrm{MC}$ (as with all the other parameters). We see that our calculated $\tau$ depends only very weakly on $h$. At first sight, this may seem surprising, given that Eqs. \eqref{eq:tauH} and \eqref{eq:tauHe} appear to be proportional to $h$. This line of reasoning would erroneously lead to the conclusion that the fractional error on $\tau$ is equal to the fractional error on $h$, which is larger than what is seen here. To understand this discrepancy, note that if we temporarily return (for the sake of simplicity) to the assumption that reionization happens instantaneously at $z_\textrm{ion}$, and further make the approximation that $\Omega_\Lambda \ll \Omega_m (1+z)^3$ for $z \approx z_\textrm{ion}$, the integrals in Eqs. \eqref{eq:tauH} and \eqref{eq:tauHe} can be evaluated analytically. The prefactor of our expression for $\tau$ then becomes $\tau \propto \Omega_b h^2(\Omega_m h^2)^{-1/2}$ to leading order. Now, recall that $\Omega_b h^2$ and $\Omega_m h^2$ are proportional to physical energy densities and hence are combinations that are directly constrained by the CMB. As a lone parameter, $h$ therefore enters only at higher order, or in the detailed astrophysics of $\overline{x_\textrm{HII} (1+\delta_b)}$ and $\overline{x_\textrm{HeIII} (1+\delta_b)}$, where its influence is much weaker. This weak dependence is welcome news in our quest to compute $\tau$, since it immunizes our estimate against possible systematic biases in $h$, such as those that are suggested by tensions between \emph{Planck}-derived values of $h$ and those determined from some supernovae measurements \cite{riess_et_al2011,bennett_et_al2014,Planck2015parameters}.

\begin{figure}[!]
	\centering
	\includegraphics[width=0.45\textwidth]{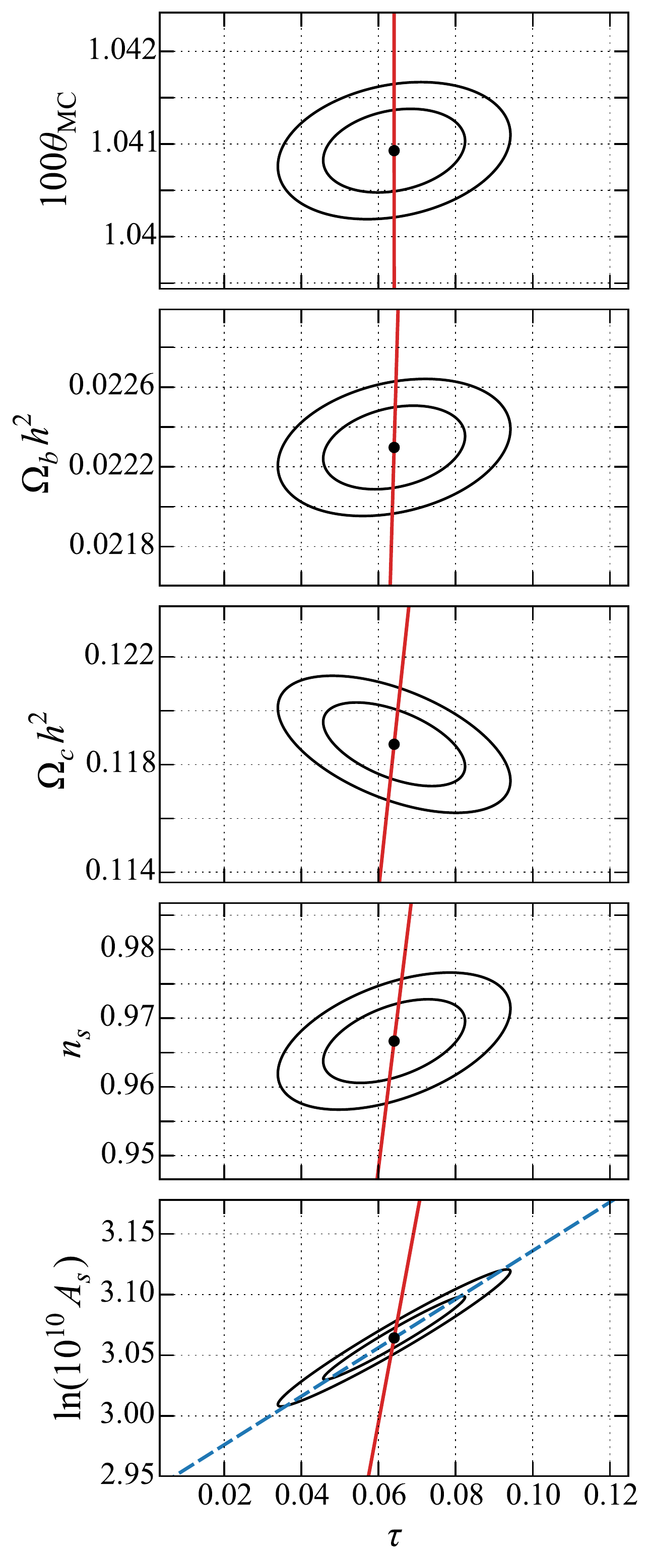}
	\caption{Likelihood contours for $\Lambda$CDM cosmological parameters as defined in the publicly released \emph{Planck} TT,TE,EE + lowP + lensing + ext dataset. Black ellipses show $68\%$ and $95\%$ confidence regions for every parameter against $\tau$. Red lines indicate values of $\tau$ as predicted in {\tt 21cmFAST} and approximated by Eq. \eqref{eq:TTTEEE_linearTau}, holding all other parameters fixed. The blue dashed line in the $\tau$-$\ln(10^{10}A_s)$ plot indicates constant $A_s e^{-2\tau}$, illustrating the strong degeneracy inherent in CMB observations that we expect to be broken by $21\,\textrm{cm}$ observations.}
	\label{fig:degenBreaking}
\end{figure}

Ultimately, the goal of a $21\,\textrm{cm}$-derived $\tau$ is not the measurement of $\tau$ itself, but rather, its elimination as a nuisance parameter for cosmological parameter estimation. With our linear relations for $\tau_\textrm{sim}$, we have tight constraints between $\tau$ and other cosmological parameters, considerably sharpening the likelihood function. To obtain some intuition for how a $21\,\textrm{cm}$-derived estimate of $\tau$ may reduce error bars, consider Fig. \ref{fig:degenBreaking}. There we plot a set of pairwise likelihood contours from \emph{Planck's} TT,TE,EE + lowP + lensing + ext dataset, pairing $\tau$ with each of the other cosmological parameters and marginalizing over all other parameters. Overlaid in red are the constraints between each parameter and $\tau$ imposed by Eq. \eqref{eq:TTTEEE_linearTau}, assuming all other parameters are fixed at their fiducial values. Crudely speaking, once a $21\,\textrm{cm}$-derived $\tau$ is folded into one's data analysis, parameter constraints must lie on the slices defined by the red lines, with a small allowance for the fact that uncertainties on the other parameters will cause the lines to become slightly blurry. One sees that in most cases there will be a non-negligible, though small, decrease cosmological parameter errors. The major exception to this is $A_s$. CMB temperature data alone have a strong degeneracy between $A_s$ and $\tau$, and remain largely unchanged if the combination $A_s e^{-2 \tau}$ is kept constant. Polarization and lensing data break this degeneracy to some extent, but there remains some residual effect (as illustrated in Fig. \ref{fig:degenBreaking} by the alignment between the ellipses and the blue line, which is a contour of constant $A_s e^{-2 \tau}$). Though $A_s$ and $\tau$ are also positively correlated in reionization simulations (essentially because larger primordial fluctuations lead to earlier structure formation, and hence earlier reionization and higher $\tau$), the slope of the relation is rather different. We may thus expect errors in $A_s$ to be considerably suppressed by the introduction of $21\,\textrm{cm}$ data.

To be more quantitative, we may incorporate a $21\,\textrm{cm}$-estimated $\tau$ into our constraints on the parameter set $\mathbf{p}$ by performing a constrained marginalization over $\tau$ to obtain a likelihood function $\mathcal{L} (\mathbf{p})$. In other words, our final likelihood is given by
\begin{equation}
\label{eq:ConstrainedOptimization}
\mathcal{L} (\mathbf{p}) = \int d\tau \, \mathcal{L}_\textrm{expt} ( \mathbf{p}, \tau) \, \delta^D \!\!\left( \tau_\textrm{sim} (\mathbf{p}) - \tau \right),
\end{equation}
where $\mathcal{L}_\textrm{expt}$ is the likelihood function of parameters from experiments alone (without the extra information imposed by our self-consistent simulations). In this paper, we will limit ourselves to considering CMB and $21\,\textrm{cm}$ experiments, although in principle, other probes of reionization such as Lyman-alpha observations can be folded into $\mathcal{L}_\textrm{expt}$. Note that without simulations, the $21\,\textrm{cm}$ power spectrum measurements place no direct constraints on $\tau$. The appearance of $\tau$ as an argument of $\mathcal{L}_\textrm{expt}$ is thus purely due to CMB experiments. The Dirac delta function term ties simulations and observations together, requiring that the inferred value for $\tau$ is consistent with the one predicted by inputting all the other cosmological parameters into simulations. If desired, modeling/simulation uncertainty may be incorporated by widening the delta function into some function of finite width (e.g., a Gaussian), although for simplicity we leave this for future work.

We note that the formalism here is a departure from the picture we have painted thus far in the paper. Until now, we have thought of $\tau$ as a parameter to be first determined by $21\,\textrm{cm}$ measurements, and then fed into CMB data analyses to refine constraints on other cosmological parameters. While conceptually tidy, this approach misses the fact that once the errors on other parameters have been brought down, the uncertainties on $\tau$ itself can be reduced once more, since Eqs. \eqref{eq:TTlowP_linearTau} and \eqref{eq:TTTEEE_linearTau} exhibit a non-negligible dependence on cosmological parameters. To account for these complications, our method here is to self-consistently require that the CMB-measured $\tau$ match a value of $\tau$ that is predicted from $21\,\textrm{cm}$ observations. With real data, this would be enacted by performing a joint fit over CMB observations and $21\,\textrm{cm}$ observations, tied together by semi-analytic simulations.

Moving forward, we will approximate $\mathcal{L}_\textrm{expt.}$ as a correlated higher-dimensional Gaussian, which is equivalent to saying that the forecasts presented in Sec. \ref{sec:CMBresults} will be based on a Fisher matrix formalism. Under the Fisher formalism, the likelihood takes the form
\begin{eqnarray}
\mathcal{L}_\textrm{expt} ( \mathbf{p}, \tau) \propto &&\,\, \exp \bigg{[}-\frac{1}{2} \bigg{(}F_{\tau \tau} (\Delta \tau)^2  + \sum_{i \neq \tau} F_{i \tau} \Delta p_i \Delta \tau \nonumber \\
&& + \sum_{j \neq \tau} F_{\tau j} \Delta p_j \Delta \tau  + \sum_{ij \neq \tau}F_{ij} \Delta p_i \Delta p_j \bigg{ )}\bigg{]}, \qquad
\end{eqnarray}
where $\Delta p_i$ and $ \Delta \tau$ are the deviations of $i$th parameter and $\tau$ about their fiducial values, respectively, and coefficients such as $F_{ij}$, $F_{i\tau}$, and $F_{\tau \tau}$ are part of a Fisher matrix $\mathbf{F}$. In the Gaussian approximation, $\mathbf{F}$ is equal to the inverse covariance of $( \mathbf{p},  \tau)$, and is an additive property of two independent experiments. In our case, it may therefore be computed by summing two contributions: the inverse covariance matrices from \emph{Planck} for basic cosmological parameters, and the $21\,\textrm{cm}$ power spectrum Fisher matrices for cosmological and astrophysical parameters (calculated in Ref. \cite{liu_and_parsons2015}).

Now, evaluating the integral in Eq. \eqref{eq:ConstrainedOptimization} is tantamount to replacing $\tau$ with $\tau_\textrm{sim} (\mathbf{p})$ in this expression. Continuing with the linear approximations to $\tau_\textrm{sim} (\mathbf{p})$ that we employed above, we have
\begin{equation}
\Delta \tau = \sum_i a_i \Delta p_i,
\end{equation}
where $\{a_i \}$ are coefficients chosen to match our linearized relations, Eqs. \eqref{eq:TTlowP_linearTau} and \eqref{eq:TTTEEE_linearTau}, and substituting this into $\mathcal{L}_\textrm{expt} ( \mathbf{p}, \tau)$ yields another Gaussian likelihood for $\mathcal{L} (\mathbf{p})$, but with modified Fisher matrix elements $F^\prime_{ij}$ given by
\begin{equation}
\label{eq:newFisher}
F^\prime_{ij} = F_{ij} + a_i F_{j \tau} + a_j F_{i \tau} + a_i a_j F_{\tau \tau}.
\end{equation}
Once this modified Fisher matrix has been obtained, it can be manipulated in the usual manner to obtain projected uncertainties on parameters. This expression will form the basis of our predictions in the following section, where we forecast the improvement in cosmological constraints from combining \emph{Planck} results with upcoming $21\,\textrm{cm}$ power spectrum measurements from HERA.

\section{Cosmological parameters with a $21\,\textrm{cm}$-derived $\tau$ constraint}
\label{sec:CMBresults}

Having established intuition and a formalism for reducing cosmological parameter uncertainties via $21\,\textrm{cm}$-derived constraints on $\tau$, we now provide some quantitative forecasts. For each of our two selected \emph{Planck} data sets, we add their inverse covariance matrices to tailored $21\,\textrm{cm}$ power spectrum Fisher matrices. These matrices are tailored in the sense that they are centered on different fiducial parameters, chosen so that when input into the {\tt 21cmFAST} simulations, the predicted values of $\tau$ match the best-fit values from the relevant \emph{Planck} datasets. We then evaluate Eq. \eqref{eq:newFisher} using either Eq. \eqref{eq:TTlowP_linearTau} or Eq. \eqref{eq:TTTEEE_linearTau}, giving final Fisher matrices that we invert to obtain final covariance matrices. As in the previous section, we assume that observations are used to form power spectra at intervals of $\Delta z = 0.3$ from $z = 6.1$ to $9.1$ inclusive for \emph{Planck} TT,TE,EE+lowP+lensing+ext, and $\Delta z = 0.5$ from $z=7.5$ to $10.5$ inclusive for \emph{Planck} TT+lowP.

Table \ref{tab:AstroParams} lists the marginalized $68\%$ limits on the astrophysical parameters that describe reionization in our model, showing the error bars that can be expected from combining \emph{Planck} priors on cosmological parameters with $21\,\textrm{cm}$ power spectrum measurements, as well as those from additionally requiring self-consistency between the CMB-measured $\tau$ and a $21\,\textrm{cm}$-informed estimate from semi-analytic simulations. Comparing the two sets of error estimates, one sees that as far as astrophysical parameters are concerned, there is little to be gained from the consistency constraint. This is to be expected from our earlier discussion of Fig. \ref{fig:21cmDegen_wTau}, where we saw that the alignment of parameter degeneracy directions meant that incorporating $\tau$ was unlikely to improve one's astrophysical parameter constraints.

\begin{table}
\caption{\label{tab:AstroParams} Fiducial values and marginalized $68\%$ confidence intervals for astrophysical parameters, within reionization scenarios tuned to fit the \emph{Planck} TT+lowP and TT,TE,EE + lowP + lensing + ext datasets. In each case, astrophysical and cosmological parameters were constrained simultaneously, with \emph{Planck} results imposed as a prior on the latter. The ``Errors from $P_{21}(k)$" are reproduced from Ref. \cite{liu_and_parsons2015} and constitute forecasted errors from HERA using power spectrum measurements only. The final column shows the errors that result from also requiring that the parameters self-consistently reproduce $\tau$ in semi-analytic simulations. Imposing self-consistency in $\tau$ has a negligible effect on astrophysical parameters, as one expects from Fig. \ref{fig:21cmDegen_wTau}.}
\begin{ruledtabular}
\begin{tabular}{llll}
\multicolumn{4}{c}{   \textbf{\emph{Planck} TT+lowP priors} }  \\
& Fid. value & Errors from  $P_{21} (k)$ & $+21\,\textrm{cm}$ $\tau$\\
\hline
$T_\textrm{vir}\left[ \textrm{K} \right]$  \dotfill & $40000$ & $\pm 7500$ & $\pm 7500$ \\
 $R_\textrm{mfp}\left[ \textrm{Mpc} \right]$ \dots \dotfill & $35.0$ & $\pm1.2$ & $\pm1.2$ \\
$\zeta$ \dotfill& $40.0$ & $\pm 4.6$ & $\pm 4.2$ \\
\hline 
\multicolumn{4}{c}{    \textbf{\emph{Planck} TT,TE,EE + lowP + lensing + ext priors} }  \\
& Fid. value & Errors from  $P_{21} (k)$  & $+21\,\textrm{cm}$ $\tau$\\
\hline
$T_\textrm{vir}\left[ \textrm{K} \right]$  \dotfill & $60000$ & $\pm 6700$ & $\pm 6600$ \\
$R_\textrm{mfp}\left[ \textrm{Mpc} \right]$ \dots \dotfill & $35.0$ & $\pm 0.82$ & $\pm 0.82$ \\
$\zeta$ \dotfill& $30.0$ & $\pm 2.0$ & $\pm 1.9$\\
\end{tabular}
\end{ruledtabular}
\end{table}

In contrast, Table \ref{tab:CosmoParams} shows that there are some improvements to cosmological parameters. While some parameters ($100 \theta_\textrm{MC}$ being the best example) are already known to such precision with \emph{Planck} that the addition of $21\,\textrm{cm}$ information does little to reduce errors, others do show improvement. In general, a better performance is obtained when the fiducial parameters are chosen to match the \emph{Planck} TT,TE,EE + lowP + lensing + ext dataset than when they are based on the \emph{Planck} TT+lowP dataset. For example, by adding $21\,\textrm{cm}$ power spectrum measurements and our $\tau$ self-consistency constraint to \emph{Planck} priors, the former dataset sees a $\sim 15\%$ reduction in errors on $\Omega_\Lambda$ and $\Omega_m$, whereas there is negligible improvement with the latter dataset. This is because the \emph{Planck} TT,TE,EE + lowP + lensing + ext parameters imply a lower redshift of reionization, which shifts the most non-trivial features in the evolution of the $21\,\textrm{cm}$ power spectrum to higher frequencies. There, both foregrounds and instrumental noise are smaller in amplitude, allowing high-significance measurements of the power spectrum that are more effective at breaking parameter degeneracies. The Hubble parameter $H_0$ stands as an exception to this general trend, with \emph{Planck} TT+lowP showing a larger error reduction. However, \emph{Planck} TT,TE,EE+lowP+lensing+ext still has smaller final error bars, having started with a more precise estimate of $H_0$.

\begin{table*}
\caption{\label{tab:CosmoParams} Fiducial values and marginalized $68\%$ confidence intervals for cosmological parameters in $\Lambda$CDM, within reionization scenarios tuned to fit the \emph{Planck} TT+lowP and TT,TE,EE+lowP+lensing+ext datasets. The ``Errors" columns show error bars using \emph{Planck} data only, ``$+P_{21} (k)$" includes $21\,\textrm{cm}$ power spectrum information (reproduced from Ref. \cite{liu_and_parsons2015}), and ``$+21\,\textrm{cm}$ $\tau$" also requires self-consistency between the CMB-measured $\tau$ and the $21\,\textrm{cm}$-predicted $\tau$. The $21\,\textrm{cm}$ observations are assumed to come from HERA. Boldfaced entries represent substantial reductions in error (arbitrarily defined as a halving or more of error bars) compared to using \emph{Planck} data only.}
\begin{ruledtabular}
\begin{tabular}{lllllcllll}
 & \multicolumn{4}{c}{\textbf{\emph{Planck} TT + lowP}} && \multicolumn{4}{l}{\textbf{ \emph{Planck} TT,TE,EE+lowP+lensing+ext}  } \\
 & Best fit & Errors &  $+P_{21} (k)$ &$+21\,\textrm{cm}$ $\tau$&& Best fit & Errors &  $+P_{21} (k)$ &  $+21\,\textrm{cm}$ $\tau$\\
\hline
\multicolumn{7}{l}{Measured parameters} \\
$\Omega_b h^2$ \dotfill & $0.02222 $&$ \pm 0.00023$ & $\pm 0.00021$ &  $\pm 0.00020$ && $0.02230 $&$\pm 0.00014$ & $\pm 0.00013$ &  $\pm 0.00013$ \\
$\Omega_c h^2$ \dotfill & $0.1197$&$ \pm 0.0022$  & $\pm 0.0021$ &  $\pm 0.0018$ && $0.1188$&$ \pm 0.0010$ & $\pm 0.00096$ &  $\pm 0.00087$ \\
$100 \theta_\textrm{MC}$\dotfill  & $1.04085 $&$\pm 0.00046$ & $\pm 0.00046$ &  $\pm 0.00045$ & &$1.04093 $&$\pm 0.00030$ & $\pm 0.00029$ & $\pm 0.00029$ \\
$\ln ( 10^{10} A_s) $ \dotfill & $3.089 $&$\pm 0.036$ & $\pm 0.023$ &  $\mathbf{\pm 0.0063}$ & &$3.064$&$ \pm 0.023$ & $\pm 0.016$ & $\mathbf{\pm 0.0053}$  \\
$ n_s $\dotfill  & $ 0.9655$&$ \pm 0.0062$ & $\pm 0.0057$ &  $\pm 0.0053$ && $0.9667$&$ \pm 0.0040$ & $\pm 0.0037$ &  $\pm 0.0035$ \\
$ \tau $ \dotfill & $0.078 $&$\pm 0.019$ & $\pm 0.013$ &  ---  && $0.066 $&$\pm 0.012$ & $\pm 0.0089$ &  --- \\
\hline
\multicolumn{7}{l}{Derived parameters} \\
$ \tau $\dotfill  & --- & ---  & --- & $ \mathbf{\pm 0.0016}$ &&---  & ---  & ---& $ \mathbf{\pm 0.00083}$ \\
$H_0 \left[ \textrm{km}\,\textrm{s}^{-1}\textrm{Mpc}^{-1}\right] \dots$ \dotfill & $67.31 $&$\pm 0.96$ & $\pm0.91$ &  $\pm 0.81$ && $67.74 $&$\pm 0.46$ & $\pm 0.43$ &  $\pm 0.41$ \\
$\Omega_\Lambda$ \dotfill & $0.685 $&$\pm 0.013$ &  $ \pm 0.013$ &  $\pm 0.011$ && $0.6911$& $ \pm 0.0062$ & $\pm 0.0057$ &  $\pm 0.0053$ \\
$\Omega_m$ \dotfill & $0.315$& $ \pm 0.013$ &   $ \pm 0.013$ & $\pm 0.011$&& $0.3089 $&$\pm 0.0062$ & $\pm 0.0057$ &  $\pm 0.0053$  \\
$\sigma_8$ \dotfill & $0.829$&$ \pm 0.014$ & $\pm 0.009$ & $\mathbf{\pm 0.0067}$ && $0.8159$& $ \pm 0.0086$ & $\pm 0.0062$ &  $\mathbf{\pm 0.0036}$\\
\end{tabular}
\end{ruledtabular}
\end{table*}

In our formalism, $\tau$ is marginalized out of our set of parameters in a self-consistent manner. It is for this reason that $\tau$ appears as a measured parameter in Table \ref{tab:CosmoParams} prior to our inclusion of $21\,\textrm{cm}$ $\tau$ information, but only as a derived parameter afterwards. To estimate errors on $\tau$, one may draw random samples of our parameters $\mathbf{p}$ from the final likelihood $\mathcal L (\mathbf{p})$ given by Eq. \eqref{eq:ConstrainedOptimization}. Samples of $\tau$ may then be obtained by inserting these randomly drawn parameters into our linearized relations for $\tau$, Eqs. \eqref{eq:TTlowP_linearTau} and \eqref{eq:TTTEEE_linearTau}, and uncertainties on $\tau$ can be estimated by examining the spread of these samples. For \emph{Planck} TT+lowP, the $1\sigma$ error on $\tau$ is $\pm 0.0015$, representing a $2\%$ measurement. Now, suppose we artificially fix the astrophysical parameters in our drawing of $\mathbf{p}$. Our error on $\tau$ then drops to $\pm 0.00055$. Note that this represents a fractional error of $0.7\%$, which is smaller than the $1.4\%$ predicted in Sec. \ref{sec:CosmoParamUncertainties}. This occurs because there are degeneracies between the cosmological parameters and the astrophysical parameters, and thus fixing the latter improves the former. If we fix the cosmological parameters and allow the astrophysical parameters to vary, the $\tau$ error is $\pm 0.00054$, almost equal to the error from only varying cosmological parameters. The patterns for the \emph{Planck} TT,TE,EE + lowP + lensing + ext dataset are similar: varying only the cosmological parameters yields an error of $\pm 0.00028$; varying only astrophysics gives $\pm 0.00036$; and varying everything gives $\pm 0.00083$, which is a $1\%$ measurement. These results confirm our intuition that once $21\,\textrm{cm}$ data are introduced, astrophysical parameter uncertainties become small enough that cosmological parameter errors must be jointly included in one's errors analysis.

\begin{figure}[!]
\centering
\includegraphics[width=0.5\textwidth]{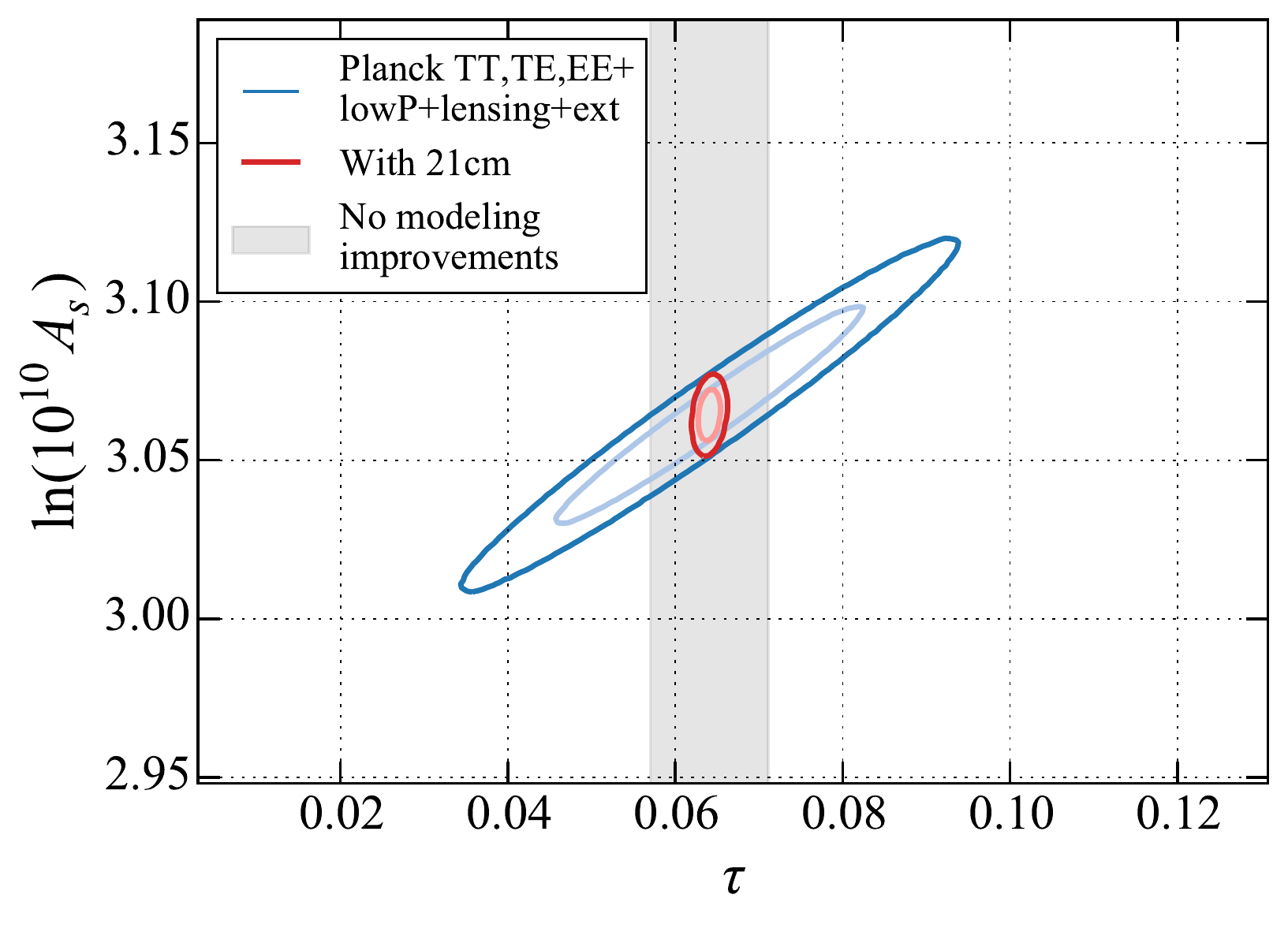}
\caption{Likelihood contours on the $\tau$-$\ln (10^{10} A_s)$ plane, with bold lines signifying $95\%$ confidence regions and light lines signifying $68\%$ confidence regions. Blue contours denote the constraints using \emph{Planck} TT,TE,EE + lowP + lensing + ext data only, while the red contours show the effect of adding $21\,\textrm{cm}$ power spectrum and---crucially---self-consistency between the CMB-measured and $21\,\textrm{cm}$-predicted $\tau$. The $21\,\textrm{cm}$ observations break the CMB degeneracy between $A_s$ and $\tau$, enabling much better constraints on both parameters. The grey band shows a width of optical depths representative of the spread of models shown in Fig. \ref{fig:InsideOutvsOutsideIn}, and is indicative of a scenario where the ionization history is known, but the density-ionization correlation is unknown. Even in the midst of such modeling uncertainty, one sees an improvement in $A_s$ errors, although we stress that such a scenario is rather pessimistic since early $21\,\textrm{cm}$ measurements will place constraints on the correlation.}
\label{fig:AsTau_w21cm}
\end{figure}

As expected from Fig. \ref{fig:degenBreaking}, the inclusion of $21\,\textrm{cm}$ information most benefits our constraints on $A_s$, since an independent constraint on $\tau$ breaks the CMB degeneracy where any changes keeping $A_s e^{-2\tau}$ constant are difficult to detect. For both \emph{Planck} datasets, an error reduction of about a factor of four is achieved in the quantity $\ln (10^{10} A_s)$. Shown in Fig. \ref{fig:AsTau_w21cm} are the $68\%$ and $95\%$ confidence regions on the $\tau$-$\ln (10^{10} A_s)$ plane for the \emph{Planck} TT,TE,EE + lowP + lensing + ext dataset. (The results for \emph{Planck} TT+lowP are qualitatively similar). One clearly sees that the $A_s e^{-2\tau}$ degeneracy is strongly broken. For reference, the grey band indicates a range of $\tau$ values that are reflective of the spread (but not the mean) of values given in Fig. \ref{fig:InsideOutvsOutsideIn} for various models of reionization. This provides an extremely conservative sense for how modeling uncertainties could degrade constraints, and even then there is some improvement from using the CMB alone. We stress, however, that this would be a very pessimistic scenario. It essentially assumes no progress in our ability to distinguish between different topologies of reionization, whereas expectations are that $21\,\textrm{cm}$ observations will be easily able to make such distinctions \cite{choudhury_et_al2009,watkinson_and_pritchard2014}. It thus seems quite likely that incorporating $21\,\textrm{cm}$ data will result in smaller error bars on $A_s$.

To interface with the large scale structure literature, it is helpful to express the normalization of the power spectrum not in terms of $A_s$, but in terms of $\sigma_8$, the root-mean-square of matter fluctuations in $8\,h^{-1}\textrm{Mpc}$ spheres at the present day, assuming linear perturbation theory. Explicitly, this is given by
\begin{equation}
\sigma_8^2 \equiv \int_0^\infty \frac{k^2 dk}{2 \pi^2} P_m (k) \left[ \frac{3 j_1 (kR)}{kR}\right]^2,
\end{equation}
where $R= 8\,h^{-1}\textrm{Mpc}$, $P_m(k)$ is the matter power spectrum at $z=0$ in linear theory, and $j_1$ is the first order spherical Bessel function of the first kind.

\begin{figure}[!]
\centering
\includegraphics[width=0.5\textwidth]{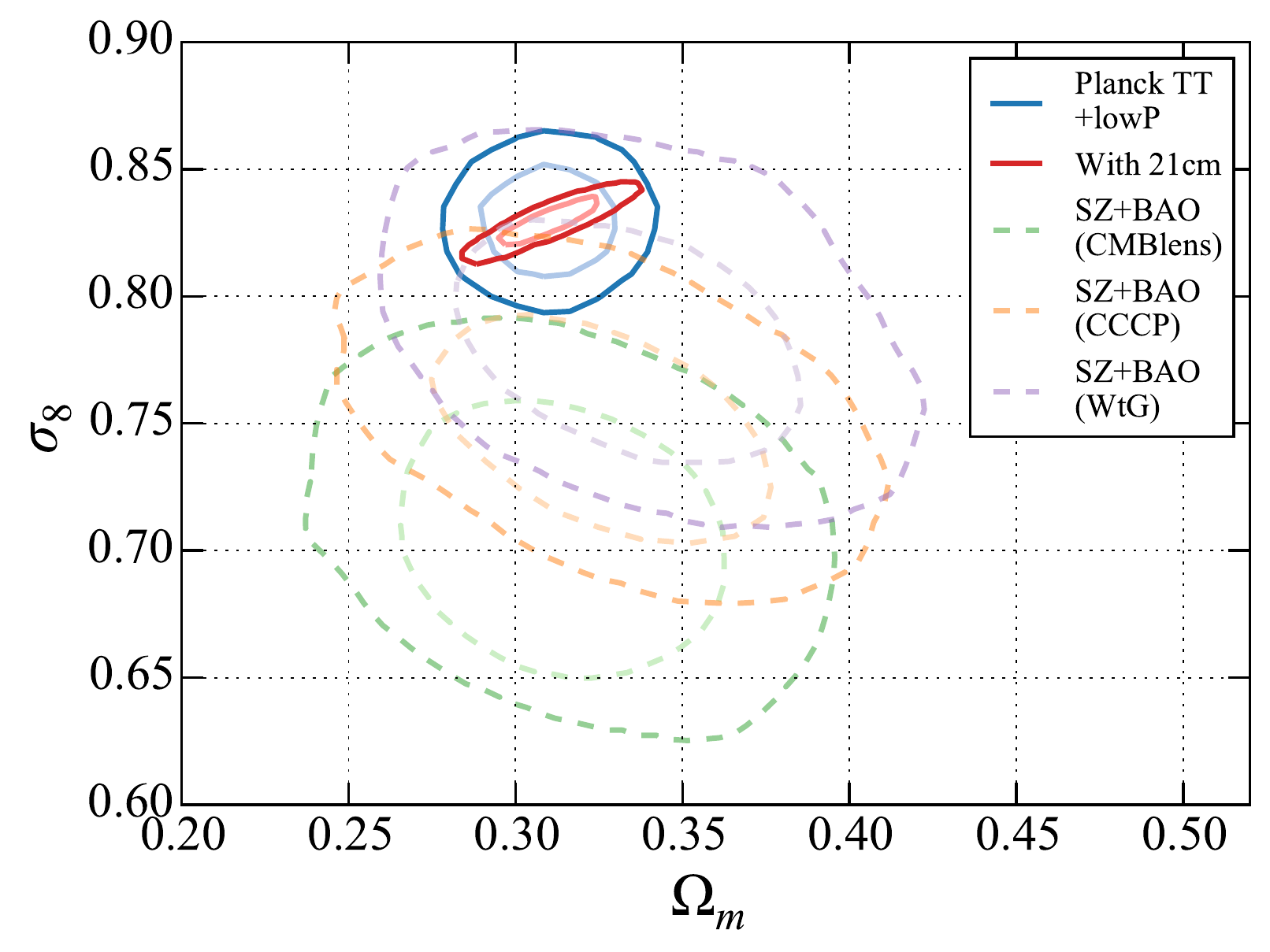}
\caption{Likelihood contours on the $\Omega_m$-$\sigma_8$ plane, with bold lines signifying $95\%$ confidence regions and light lines signifying $68\%$ confidence regions. Blue contours denote the constraints using \emph{Planck} TT+lowP data only, while red incorporates $21\,\textrm{cm}$ power spectrum and self-consistent $\tau$ information. Dashed contours denote constraints from combining \emph{Planck's} SZ cluster counts, BAO, and BBN (as published in Ref. \cite{Planck2015clusters}). In green are constraints using a CMB lensing-calibrated prior on the cluster mass bias factor (CMBlens). In orange and purple are constraints based on calibrations using gravitational shear mass measurements from the Canadian Cluster Comparison Project (CCCP) and Weighing the Giants (WtG) program, respectively.}
\label{fig:omegamsigma8}
\end{figure}

In Fig. \ref{fig:omegamsigma8}, we translate our parameter constraints into constraints on $\Omega_m$ and $\sigma_8$. Shown in blue solid lines are the $68\%$ and $95\%$ likelihood contours from the original \emph{Planck} TT+lowP dataset, while the solid red contours show the improvement from adding $P_{21} (k)$ and our $\tau$ constraints. In addition to a general shrinking of the errors, one also sees a reorientation of the likelihood contours. Whereas the errors in $\sigma_8$ and $\Omega_m$ are largely independent of each other using \emph{Planck} data alone, this is not the case once $21\,\textrm{cm}$ information is included. To understand why this occurs, recall that computing $\sigma_8$ requires integrating the \emph{present-day} matter power spectrum. It thus depends not only on the primordial fluctuation amplitude $A_s$, but also on the evolution of perturbations, which depends on parameters such as $\Omega_m$. In the case of the \emph{Planck} data alone, the error on $\sigma_8$ is dominated by the error on $A_s$, which masks the dependence on other parameters. With $21\,\textrm{cm}$ $\tau$ constraints, the errors on $A_s$ are reduced by so much that they no longer drive the errors on $\sigma_8$. Instead, uncertainties in perturbation growth become the dominant source of error, leading to correlations between $\sigma_8$ and $\Omega_m$.
%

Sharper constraints on $\sigma_8$ have the potential to shed light on current tensions between cosmological constraints derived from the primary CMB and those that are derived from galaxy cluster measurements combined with BAO and BBN \cite{Planck2015clusters}. Aside from our $\Omega_m$-$\sigma_8$ projections, Fig. \ref{fig:omegamsigma8} also shows likelihood contours from galaxy cluster counts of Sunyaev-Zeldovich (SZ) clusters (reproduced from Ref. \cite{Planck2015clusters}) using various calibration methods for the mass bias. Orange and purple contours are constraints from gravitational shear-based calibration methods using data from the Canadian Cluster Comparison Project (CCCP) and the Weighing the Giants (WtG) program, respectively. The green contours use a CMB lensing-based calibration for the mass bias. Here we focus exclusively on the \emph{Planck} TT+lowP dataset in an effort to separate the high redshift constraints on the CMB from the low redshift constraints from clusters. Moderate tensions are visually present at varying degrees depending on the SZ calibration method. In tandem with further SZ mass calibration studies, the reduced errors on $\sigma_8$ from the addition of $21\,\textrm{cm}$ $\tau$ constraints have the potential to either resolve or sharpen tensions. If tensions remain, their increased statistical significance would hint at the existence of systematics or the need for an extension to \emph{Planck's} six-parameter base model.

As an example of an extended cosmological model, consider a non-zero neutrino mass. Massive neutrinos alter the kinematics of our Universe's expansion \cite{pan_and_knox2015}. Additionally, neutrinos dampen structure growth on scales finer than their free-streaming length \cite{hu_and_eisenstein1998,eisenstein_and_hu1999,hu_et_al1998}, leading to deficits in power on small scales that are more pronounced if the sum of the neutrino masses $\sum m_\nu$ is large. Upcoming precision measurements targeting both the kinematic and structure growth signatures may therefore yield a detection of a non-zero neutrino mass. Among other studies, Ref. \cite{allison_et_al2015} provides forecasts for the expected performance for the combination of DESI and ``Stage 4" (S4) CMB experiments \cite{wu_et_al2014,Abazajian_et_al2015}. That work assumes DESI measurements of the BAO signature and CMB measurements of lensed TT, TE, and EE power spectra, along with a measurement of the CMB convergence power spectrum. A crucial parameter in the forecasting exercise is the minimum multipole $\ell_\textrm{min}$ that is assumed to be recoverable in an S4 measurement. Ref. \cite{allison_et_al2015} found that with $\ell_\textrm{min} =  5$, S4 experiments could constrain $\sum m_\nu$ to $\pm 15 \,\textrm{meV}$ ($68\%$ confidence) when analyzed in conjunction with \emph{Planck} polarization and DESI data; with $\ell_\textrm{min} =  50$, the error degrades to $\pm 19 \,\textrm{meV}$. Now, neutrino oscillation measurements constrain $\sum m_\nu$ to have a \emph{minimum} value of $60\,\textrm{meV}$ \cite{smirnov2006,fogli_et_al2012,feldman_et_al2013,Abazajian_et_al2015}. If we take this to be our fiducial value for $\sum m_\nu$, going from $\ell_\textrm{min} =  5$ to $\ell_\textrm{min} =  50$ then represents a degradation from a $\sim\!4 \sigma$ to a $\sim\!3 \sigma$ detection. At this point, it is not clear what $\ell_\textrm{min}$ would be for ground-based S4 experiments, due to difficulties with atmospheric and ground contamination \cite{allison_et_al2015}, and it is likely that the best measurements will come from a combination of ground-, balloon-, and space-based experiments, particularly when foregrounds are taken into account \cite{errard_et_al2015}.

\begin{figure}[!]
\centering
\includegraphics[width=0.5\textwidth]{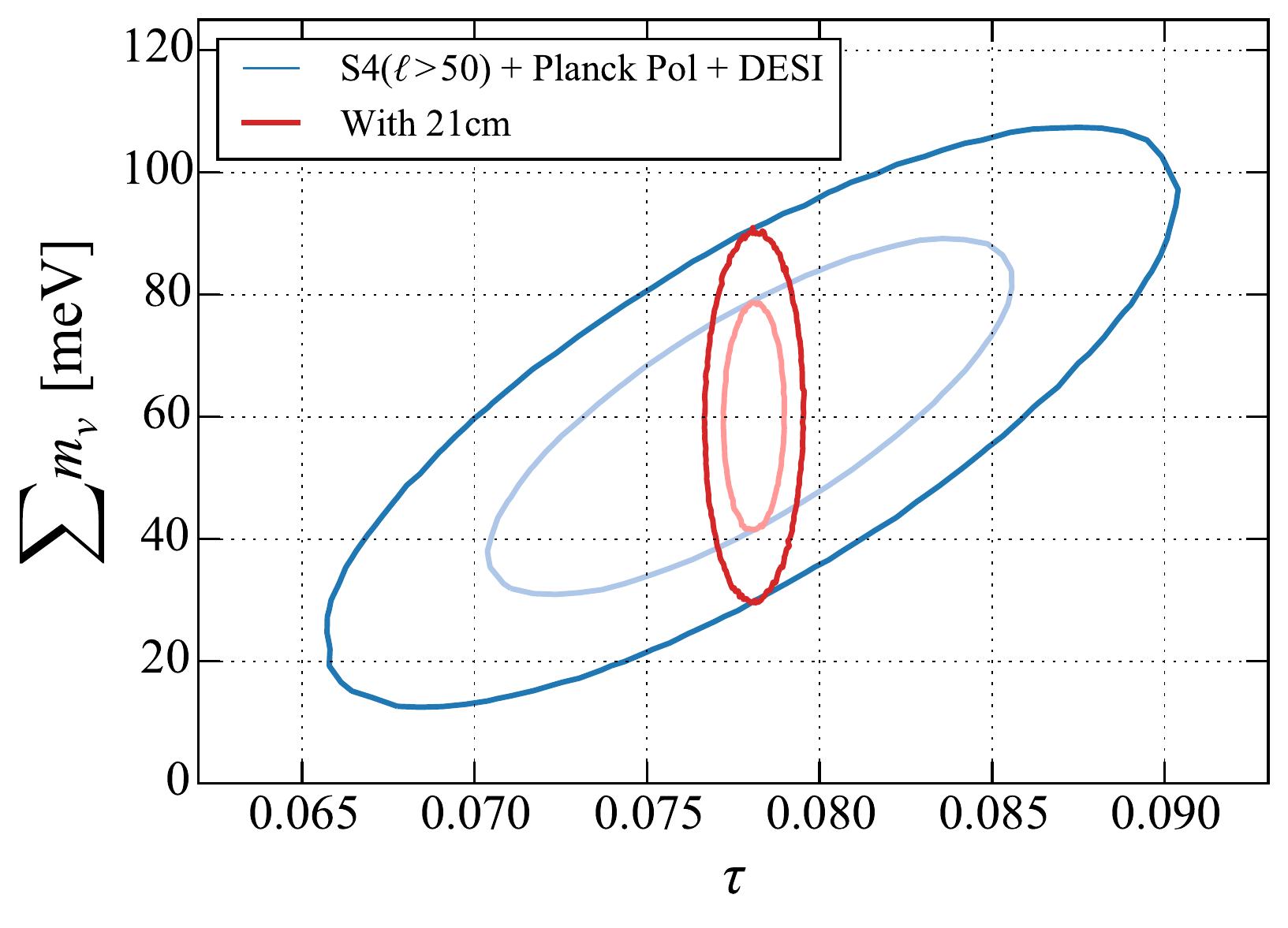}
\caption{Likelihood contours on the $\tau$-$\sum m_\nu$ plane, with bold lines signifying $95\%$ confidence regions and light lines signifying $68\%$ confidence regions. Blue contours denote the constraints using S4($\ell > 50$) + \emph{Planck} polarization + DESI data only, while the red contours show the effect of adding $21\,\textrm{cm}$ power spectrum and self-consistency between the CMB-measured and $21\,\textrm{cm}$-predicted $\tau$. Early $21\,\textrm{cm}$ observations will confirm models of reionization, allowing high sensitivity measurements to predict $\tau$. This will break the CMB degeneracy between $\tau$ and $\sum m_\nu$ and enable improved constraints on the neutrino mass.}
\label{fig:SumnuTau_w21cm}
\end{figure}

That the neutrino mass forecasts depend on $\ell_\textrm{min}$ is largely due to a degeneracy between $\tau$ and $\sum m_\nu$. This arises because neutrinos suppress structure on small scales, which can be mimicked by a lower $A_s$. As described above, this is in turn degenerate with $\tau$, leading to a $\tau$-$\sum m_\nu$ degeneracy. Accessing the lowest $\ell$ modes enable S4 experiments to make precise measurements of the reionization bump signature discussed in Sec. \ref{sec:Intro}, breaking the $\tau$-$A_s$ degeneracy (which we saw in Fig. \ref{fig:AsTau_w21cm} still exists with current \emph{Planck} data). Higher $\ell_\textrm{min}$ values for S4 experiments compromise their ability to do this degeneracy breaking. With the $21\,\textrm{cm}$ line, however, we recover this ability. Using Fisher matrices from Ref. \cite{allison_et_al2015}, we use the formalism of Sec. \ref{eq:Pkformalism} to again predict the effect of self-consistently including $21\,\textrm{cm}$ information. Fig. \ref{fig:SumnuTau_w21cm} illustrates how this breaks the $\tau$-$\sum m_\nu$ degeneracy, with the blue contours showing the constraints from S4($\ell > 50$) + \emph{Planck} polarization + DESI and the red contours additionally incorporating $21\,\textrm{cm}$ information from HERA. The fiducial value for $\sum m_\nu$ is taken to be the minimal $60\,\textrm{meV}$; the fiducial value for $\tau$ is taken to be $0.078$. This is the best-fit $\tau$ value for the \emph{Planck} TT+lowP dataset, so we use the $21\,\textrm{cm}$ Fisher matrix that is matched to \emph{Planck} TT+lowP parameters, but in practice we find that the results are essentially the same assuming \emph{Planck} TT,TE,EE + lowP + lensing + ext. We see from Fig. \ref{fig:SumnuTau_w21cm} that the $\tau$-$\sum m_\nu$ degeneracy is broken, with the error on 
$\sum m_\nu$ reduced from $\pm 19 \,\textrm{meV}$ to $\pm 12 \,\textrm{meV}$. (Forecasted errors\footnote{Importantly, note that with S4 and DESI cosmological parameters, the fractional errors on a $21\,\textrm{cm}$-predicted $\tau$ are comparable to the uncertainties from $\tau$ due to helium reionization, as predicted in Sec. \ref{sec:astroUncertainties}. We have thus implicitly assumed that by the time S4 and DESI data are available, current probes of helium reionization will have improved astrophysical models sufficiently to enable tight predictions of $\tau_\textrm{He}$.} on all cosmological parameters are given in Table \ref{tab:S4CosmoParams}). This demonstrates that even if S4 experiments are unable to precisely constrain the reionization bump, $21\,\textrm{cm}$ cosmology can fill in the missing information. With an error of $12 \,\textrm{meV}$, a cosmological determination of the neutrino mass becomes a $5\sigma$ detection even with the most pessimistic fiducial value of $\sum m_\nu = 60\,\textrm{meV}$.

\begin{table}
\caption{\label{tab:S4CosmoParams} Fiducial values and $68\%$ confidence limits on a $\Lambda$CDM plus neutrino mass ($\sum m_\nu$) model, within a reionization model tuned to fit the \emph{Planck} TT+lowP dataset. The errors are computed first assuming a Stage 4 CMB experiment able to access multipoles down to $\ell_\textrm{min} = 50$, analyzed in conjunction with DESI and \emph{Planck} polarization data (``$S4_{\ell < 50}$+DESI+\emph{Planck} Pol"). These datasets are then supplemented with HERA measurements of the $21\,\textrm{cm}$ power spectrum and self-consistent reionization simulations. The addition of $21\,\textrm{cm}$ information reduces error bars on $\sum m_\nu$ and allows a $5\sigma$ detection of the neutrino mass even if $\sum m_\nu$ is at its minimum value of $60\,\textrm{meV}$ allowed by neutrino oscillation experiments.}
\begin{ruledtabular}
\begin{tabular}{llll}
 & Fiducial & S4$_{\ell > 50}$+DESI & $+P_{21} (k)$ \\
Parameter & Value & +\emph{Planck} Pol& $+21\,\textrm{cm}$ $\tau$ \\
\hline
$\Omega_b h^2$ \dotfill & 0.0222 & $\pm 0.00003$ & $\pm 0.00003$ \\
$\Omega_c h^2$ \dotfill  & 0.1197 &$\pm 0.00038$ & $\pm 0.00022$\\
$100 \theta_\textrm{MC}$ \dotfill  & 1.04085 &$\pm 0.00031 $ & $\pm 0.00022 $\\
$\ln ( 10^{10} A_s) $ \dotfill  & 3.089 & $\pm0.0091 $ & $\pm0.0016 $\\
$n_s$ \dotfill  & 0.9655 & $\pm 0.0017 $ & $\pm 0.0015$\\
$\tau$ \dotfill  & 0.078  & $\pm 0.005 $ & $\pm 0.00058$ \\
$\sum m_\nu$ [meV] \dots & 60 & $\pm 19$ & $\pm12 $\\
\end{tabular}
\end{ruledtabular}
\end{table}

Though the predictions in this section have been primarily based on HERA, our qualitative conclusions should hold for any next-generation high signal-to-noise $21\,\textrm{cm}$ experiment. For example, consider the SKA's constraining power under the base $\Lambda$CDM model (i.e., without fitting for the neutrino mass). Rerunning our computations using the ``halved dipoles per station" SKA configurations\footnote{The ``halving" is with respect to the original SKA design, and is a consequence of the recent re-budgeting process within the SKA collaboration.} presented in Ref. \cite{greig_et_al2015b}, we obtain an error of $\pm 0.00060$ for $\tau$ and an error of $\pm 0.0052$ for $\ln (10^{10} A_s)$ using the \emph{Planck} TT,TE,EE + lowP + lensing + ext dataset. With \emph{Planck's} TT+lowP dataset, we obtain a $\pm 0.0013$ error on $\tau$ and a $\pm 0.0060$ error on $\ln (10^{10} A_s)$. (In all cases, the quoted errors refer to $68\%$ confidence). Comparing these numbers to the HERA results in Table \ref{tab:CosmoParams}, we see that the SKA delivers small improvements in precision, but not at a level that results in qualitatively new science.  We find this to be true whether we use the ``standard" SKA baseline configuration of Ref. \cite{greig_et_al2015b} or their ``compact" configuration. There are a number of reasons for the lack of significant improvement beyond HERA with the SKA.\footnote{Note that this does not preclude the possibility of precision cosmology with the SKA using methods beyond the formalism of this paper. Our claim here only encompasses improvements in cosmological parameters that arise from a better $\tau$ measurement, and does not include the extraction of other cosmological information that the SKA might provide.} First, recall from Fig. \ref{fig:21cmDegen_wTau} that astrophysical degeneracies from $21\,\textrm{cm}$ power spectrum measurements are mostly aligned with contours of constant $\tau$. For the most part, this is a helpful feature, as it decreases (though does not eliminate) the exposure of our $\tau$ estimate to astrophysical uncertainties. However, this also means that once a high signal-to-noise measurement of the $21\,\textrm{cm}$ power spectrum is achieved, one is in a regime of diminishing returns. Greater collecting area does result in the shrinking of the ellipses in Fig. \ref{fig:21cmDegen_wTau}, but the degeneracy direction is already so well aligned with the $\tau$ contours that the decrease in errors on $\tau$ is not as large as one might hope for. Additionally, the effect of a more precise, SKA-based $\tau$ on $A_s$ is rather small. This is again due to diminishing returns. From Fig. \ref{fig:AsTau_w21cm}, one sees that a HERA-based $\tau$ has (for all intents and purposes) broken the CMB $\tau$-$A_s$ degeneracy. Even greater precision on $\tau$ thus does little to further improve constraints on cosmological parameters.

\section{Model dependence of $\tau$ constraints from $21\,\textrm{cm}$ power spectrum measurements}
\label{sec:ModelDependence}

In the previous section, we saw how $21\,\textrm{cm}$ power spectrum measurements could be used in conjunction with semi-analytic simulations of reionization to place stringent constraints on $\tau$, considerably reducing errors on cosmological parameters in the process. While powerful, the danger in such an approach is that it is rather model-dependent, and requires that the semi-analytic simulations correctly capture the essential features of reionization. In this section, we discuss some of the potential problems associated with model-dependent constraints, and how they may be alleviated.

Of greatest concern is the extrapolation that is assumed in this paper, where semi-analytic simulations are used to extrapolate ionization histories to high redshifts.\footnote{In this section only, ``high redshift" refers to $z \gtrsim 13$, and ``low redshift" refers to $z \lesssim 13$.} In a way, this is worrying because there are no direct measurements of the highest redshifts, where different physical effects may come into play. This risk may be mitigated in several ways. First, we note that even though we have concentrated on the use of $21\,\textrm{cm}$ measurements at $z < 10.5$ (where the signal to noise is highest) in this paper, most instruments can reach higher redshifts. For example, PAPER and HERA can in principle reach $z \sim 13.2$, and the MWA reaches $z \sim 16$. LOFAR probes up to $z \sim 11$ with its high-band system, and from $ z \sim 17 $ to $z \sim 22$ with its low-band system. In all these cases, the upper redshift boundary is somewhat uncertain, since signal-to-noise typically drops off towards the edge of one's band. However, suppose we optimistically assume that HERA can make meaningful observations up to the edge of its design specification (perhaps with extra integration time making up for reduced sensitivity). In that case, only redshifts at $z>13.2$ are wholly dependent on our theoretical model. In our fiducial model for \emph{Planck} TT,TE,EE+lowP+lensing+ext, the $z > 13.2$ contribution to the optical depth is $0.00062$. This is less than the $1\sigma$ error of $\pm 0.00083$ predicted for combined \emph{Planck} and $21\,\textrm{cm}$ constraints in Table \ref{tab:CosmoParams}, and comparable to the constraint of $\pm 0.00058$ listed in Table \ref{tab:S4CosmoParams} from combined Stage 4 and $21\,\textrm{cm}$ analyses. This analysis suggests that redshift regimes not probed by near-term instruments make only a small contribution to $\tau$. We find that this contribution is even smaller than we have estimated for our fiducial model when we repeat our rough calculations using ionization history curves from the state-of-the-art Cosmic Reionization on Computers simulations \cite{gnedin_et_al2014,gnedin_and_kaurov2014}.

It is important to note, however, that models with extended ionization tails beyond $z> 13$ do exist. These typically involve feedback mechanisms interacting with star formation in low-mass halos, and can result in ionized fractions as high as $\sim 10^{-1}$ at $z \gtrsim 13$ \cite{Iliev_et_al2007,Iliev_et_al2012,ahn_et_al2012,park_et_al2013}. In such scenarios, the high redshift contribution is non-negligible and must be modeled. The crucial quantity is then not the absolute size of this contribution, but the precision to which this can be modeled (note that this is also the case for the models without a significant high redshift contribution). In general, the error on the modeling will be smaller than the size effect itself, particularly since certain pieces of high-redshift physics (such as the relative importance between high-mass and low-mass galaxies in the reionization process) have consequences that can be probed by lower redshift experiments. For example, supersonic relative velocity effects between dark matter and baryons \cite{tseliakhovich_and_hirata2010} suppress the formation of low-mass minihalos \cite{dalal_et_al2010,tseliakhovich_et_al2011,fialkov_et_al2012}, resulting in observable effects in the $21\,\textrm{cm}$ power spectrum even down to $z \sim 10$ \cite{cohen_et_al2015}.

Of course, studying high redshift processes through low redshift observations will be somewhat indirect, reducing the stringency of the resulting constraints. Fortunately, uncertainties in the high redshift processes can be tolerated so long as they do not affect the ionization history, in a similar fashion to the way the degeneracies shown in Fig. \ref{fig:21cmDegen_wTau} have a relatively small effect on $\tau$. In addition, there is ongoing progress to extend the redshift range of $21\,\textrm{cm}$ experiments. For example, the HERA collaboration is currently designing an updated feed that will enable observations to $z \sim 30$. Power spectrum measurements based on such observations are forecasted to produce excellent constraints on the heating processes relevant at high redshifts \cite{ewall-wice_et_al2015}, and the potential even exists for model-independent reconstructions of the X-ray spectra responsible for this heating \cite{fialkov_et_al2015}. The SKA and the Long Wavelength Array (LWA \cite{taylor_et_al2012}) represent yet other opportunities to make power spectrum measurements (and perhaps even imaging) observations at high redshifts.

Complementary to upcoming measurements of the high redshift $21\,\textrm{cm}$ power spectrum will be global signal experiments. These experiments aim to measure the all-sky-averaged brightness temperature as a function of frequency, $\overline{\delta T}_b (\nu)$. Proposed or in-progress experiments include the Experiment to Detect the Global Epoch of Reionization Signature \citep{bowman2010}, Large-Aperture Experiment to Detect the Dark Ages \citep{greenhill2012}, Dark Ages Radio Explorer \citep{burns2012}, Broadband Instrument for Global Hydrogen Reionization Signal \cite{Sokolowski_et_al2015}, Shaped Antenna Measurement of the Background Radio Spectrum \cite{patra_et_al2013}, Zero-spacing Interferometer Measurements of the Background Radio Spectrum \cite{mahesh_et_al2014}, and Sonda Cosmol\'{o}gica de las Islas para la Detecci\'{o}n de Hidr\'{o}geno Neutro \citep{voytek2014}. Many of these are designed to reach redshifts up to $z\sim 35$ \cite{burns2012,vedantham_et_al2014,bernardi_et_al2015}, and again, are expected to make measurements of relevant high redshift physics. For instance, these experiments have the potential to place tight constraints on the Lyman-Werner radiation background \cite{mirocha_et_al2015}, which is a crucial ingredient in many of the self-regulating feedback processes that sustain the extended high redshift ionization tails seen in some simulations.

While in this paper we have focused primarily on the ability of the CMB to constrain cosmology, future experiments may also provide precise constraints on reionization that include effects from high redshift physics. Upcoming high sensitivity arcminute-scale resolution polarization experiments will make high sensitivity measurements of the kinetic Sunyaev-Zel'dovich (kSZ) effect, providing some information on the duration of reionization from CMB experiments \cite{calabrese_et_al2014,alvarez2015}. To be fair, Ref. \cite{park_et_al2013} found that kSZ measurements may not be very sensitive to self-regulation effects if made at spherical harmonic wavenumber $\ell \sim 3000$, which is the focus of current measurement attempts. However, they also find that the impact of self-regulation becomes more pronounced at higher $\ell$, a regime that may be enabled by new experiments combined with ongoing advances in multi-wavelength isolation of foregrounds and thermal Sunyaev-Zel'dovich contaminants. It should also be noted that the relative baryon-dark matter velocity effect \cite{tseliakhovich_and_hirata2010} (discussed above) modulates minihalo ionization bubbles on very large scales, and may also enhance the kSZ signal at lower $\ell$, which is an effect ignored in current kSZ simulations of this epoch.

Finally, there exists the possibility of exotic new physics that is unaccounted for even in our most sophisticated models. A prime example of this would be early ($z \sim 100$ to $200$) dark matter annihilations, which could have significant effect on $\tau$ without correspondingly large perturbations to the $21\,\textrm{cm}$ power spectrum \cite{kaurov_et_al2015b}. In this case, however, the discrepancy between our $21\,\textrm{cm}$-derived $\tau$ and the CMB-derived $\tau$ would be a welcome tension, and would only become apparent by performing the analyses proposed in this paper

In summary, though the methods described thus far are model-dependent, there exist ample opportunities to refine our models over the next few years. If the true ionization history of our Universe is close to zero beyond $z \gtrsim 13$ (as many models suggest), almost the entirety of the optical depth $\tau$ will be sourced at redshifts where direct observations will soon be available, and little extrapolation to high redshifts will be required. On the other hand, if models with long ionization tails towards high redshifts turn out to be correct, some extrapolation may be initially necessary, but ongoing progress with high redshift ($z \lesssim 35$) experiments should enable further model refinement. In any case, we expect the forecasts in this paper to be revised as our understanding of reionization becomes increasingly informed by observations, and the work here is intended only to be a proof-of-concept study to highlight the potential of using $21\,\textrm{cm}$ information to augment CMB constraints.

\section{Using global signal measurements to derive model-independent $\tau$ constraints}
\label{sec:GlobalSig}

As discussed in the previous section, the use of the $21\,\textrm{cm}$ power spectrum measurements to constrain $\tau$ requires the use of astrophysical models, which carries the risk of model-dependency. In this section, we discuss how direct observations of the $21\,\textrm{cm}$ temperature field (rather than its power spectrum) can provide direct, model-independent constraints on $\tau$.

Ignoring hard-to-measure patchy effects \cite{dvorkin_et_al2009,su_et_al2011,natarajan_et_al2013}, $\tau$ is effectively an angularly averaged quantity, with one value across the entire sky. Correspondingly, in an attempt to use direct measurements of the $21\,\textrm{cm}$ brightness temperature to constrain $\tau$, it is not necessary to measure the three-dimensional distribution $\delta T_b (\mathbf{\hat{n}}, \nu)$. Instead, the angularly averaged quantity $\overline{\delta T}_b (\nu)$ suffices. This is precisely the purview of the global signal experiments discussed in the previous section.

To see how global signal measurements can be used to constrain $\tau$, suppose that peculiar velocity contribution to the brightness temperature can be ignored (we will address it below). Further assume that the hydrogen spin temperature is much greater than the CMB temperature, $T_s \gg T_\gamma$. This approximation is expected to be justified towards the middle and end of reionization, when the spin temperature is tightly coupled to the kinetic temperature of the IGM, which is strongly heated by X-rays and/or shocks from filamentary structure or exotic mechanisms such as dark matter annihilation \cite{gnedin_and_shaver2004,pritchard_and_loeb2010,mirabel_et_al2011,valdes_et_al2013,evoli_et_al2014}. Under these assumptions, Eq. \eqref{eq:deltaTdef} simplifies to $\delta T_b(\mathbf{\hat{n}}, \nu) \approx \delta T_{b0}\, x_\textrm{HI}  (1 + \delta_b)$. Taking the angular average of this, the resulting global signal $\overline{\delta T}_b (\nu)$ and the density-weighted ionized fraction are seen to be related via a simple linear equation, namely
\begin{equation}
\label{eq:globalSigDirect}
1- \frac{\overline{\delta T}_b (\nu)}{\delta T_{b0}} = \overline{x_\textrm{HII} (1+\delta_b)}.
\end{equation}
As we saw in Sec. \ref{sec:astroUncertainties}, the density-weighted ionized fraction is the crucial quantity in an accurate estimate of $\tau$. Whereas power spectrum measurements require model-dependent simulations to infer $\overline{x_\textrm{HII} (1+\delta_b)}$, global signal measurements can do so directly in a model-independent way.

Of course, our claim of model-independence holds only if our assumption of $T_s \gg T_\gamma$ is true, since computing $T_s$ requires detailed models of the radiative backgrounds and atomic physics \cite{hirata2006,hirata_and_sigurdson2007}. The spin temperature approximation will almost certainly fail at the beginning of the reionization epoch, prior to the completion of reheating. Unfortunately, discarding this approximation requires simulating the complicated astrophysics and atomic processes that govern $T_s$, which of course require a reionization model. The best that we can do is to lower our ambitions, and to restrict our global signal constraints to the lower redshift contributions to $\tau$. Ideally, one would first use the model-dependent power spectrum methods of the previous sections to derive an overall $\tau$ constraint, which could then be checked for consistency against a model-independent estimate of the low-redshift contributions from global signal measurements.

We now consider the peculiar velocity term. This is due to redshift space distortions, where peculiar velocities mean that it is incorrect to assume that frequencies and comoving radial coordinates are simply related by the Hubble flow. The distribution of emission is thus different in frequency (or redshift) space than in real comoving spatial coordinates. However, whether one works in redshift space or real space, the total integrated emission along the radial line-of-sight is by definition the same. Such an integral is precisely what is evaluated when predicting $\tau$. The peculiar velocity term can therefore be neglected. We do note that in practice, our simulations show a small ($\lesssim 0.2\%$) difference between integrating $1- \overline{\delta T}_b (\nu) / \delta T_{b0}$ and $\overline{x_\textrm{HII} (1+\delta_b)}$. This is likely because Eq. \eqref{eq:deltaTdef} is itself an approximate treatment of redshift space distortions, strictly valid only when $\partial v_r / \partial r \ll H$. For the forecasts below we simply ignore this discrepancy (because it is small), and note that it would not appear in an actual measurement.

Having established that $1- \overline{\delta T}_b (\nu)/\delta T_{b0}$ is a good approximation to $\overline{x_\textrm{HII} (1+\delta_b)}$, we may substitute Eq. \eqref{eq:globalSigDirect} into Eq. \eqref{eq:tauH}. Limiting our computation of the optical depth to the contribution between $z=0$ and some relatively low redshift $z$ (in keeping our assumption that $T_s \gg T_\gamma$), we obtain
\begin{eqnarray}
\tau(z) = &&\frac{H_0 \Omega_b \sigma_Tc}{4 \pi G \Omega_m \mu m_p} \left[ \sqrt{\Omega_\Lambda + \Omega_m (1+z)^3} -1 \right] \nonumber \\
&&\quad - \frac{16 \sigma_T \nu_0^2 k_B}{3 \hbar c^2 A_{10}} \int_0^z dz^\prime \, \overline{\delta T}_b \sqrt{1+z^\prime},
\end{eqnarray}
where it is understood that $\overline{\delta T}_b$ is to be evaluated at frequency $\nu_{21} / (1+z^\prime)$. In deriving this expression, we used Eq. \eqref{eq:TbPrefactor} to express $\delta T_{b0}$ explicitly in terms of fundamental constants, and for the second term only, made the approximation that $\Omega_\Lambda \ll \Omega_m (1+z)^3$. This is an excellent approximation even at the high levels of precision being pursued here, since the approximation becomes bad only at the lowest redshifts, but by then reionization is complete and $\overline{\delta T}_b$ is zero.

The first term in our expression for $\tau(z)$ is the optical depth that would have resulted had our Universe been ionized throughout cosmic history. Conveniently, it takes the same form as Eq. \eqref{eq:exactConstant}, which as we argued in Sec. \ref{eq:Pkformalism}, has errors that are dominated not by the Hubble parameter, but by the much more precisely known combinations $\Omega_b h^2$ and $\Omega_m h^2$. At $z=8.5$, this term has a standard deviation of $4.7\times 10^{-4}$ for \emph{Planck} TT,TE,EE + lowP + lensing + ext and a standard deviation of $8.7 \times 10^{-4}$ for \emph{Planck} TT+lowP, with a central value of $0.063$ for both.

The second term in $\tau(z)$ is a deficit term. It quantifies the deficit in the CMB optical depth that arises because our Universe was neutral for part of its past. Importantly, we see that all cosmological parameters have canceled out of this term, leaving only fundamental constants that can be determined to high precision in a laboratory. Though this cancellation is remarkable, it is not surprising, since the $21\,\textrm{cm}$ brightness temperature is ultimately a direct measurement of the optical depth of clouds of neutral hydrogen at high redshift. This neutral hydrogen optical depth is precisely what sources the deficit in the CMB optical depth. The factor of $\sigma_T  / A_{10} $ in the prefactor of the expression acts as a conversion factor to account for the relatively small cross-section of the $21\,\textrm{cm}$ line compared to that of Thomson scattering.

With cosmological factors canceling out, the only source of error in the deficit term is thus the measurement uncertainty of the global signal $\overline{\delta T}_b$. Computing the variance $(\Delta \tau)^2$ of $\tau(z)$, we have
\begin{eqnarray}
\label{eq:DeltaTauDoubleIntegral}
(\Delta \tau)^2 = &&\left( \frac{16 \sigma_T \nu_0^2 k_B}{3 \hbar c^2 A_{10}} \right)^2 \nonumber \\
&&\times \int_0^z \! \!\int_0^z dz^\prime dz^{\prime \prime} \Sigma (z^\prime, z^{\prime \prime}) \sqrt{(1+z^\prime) (1+z^{\prime \prime})}, \qquad
\end{eqnarray}
where $\Sigma (z^\prime, z^{\prime \prime})$ is the error covariance between measurements of 
the global signal at redshift $z^\prime$ and $z^{\prime \prime}$

\begin{figure}[!]
\centering
\includegraphics[width=0.5\textwidth]{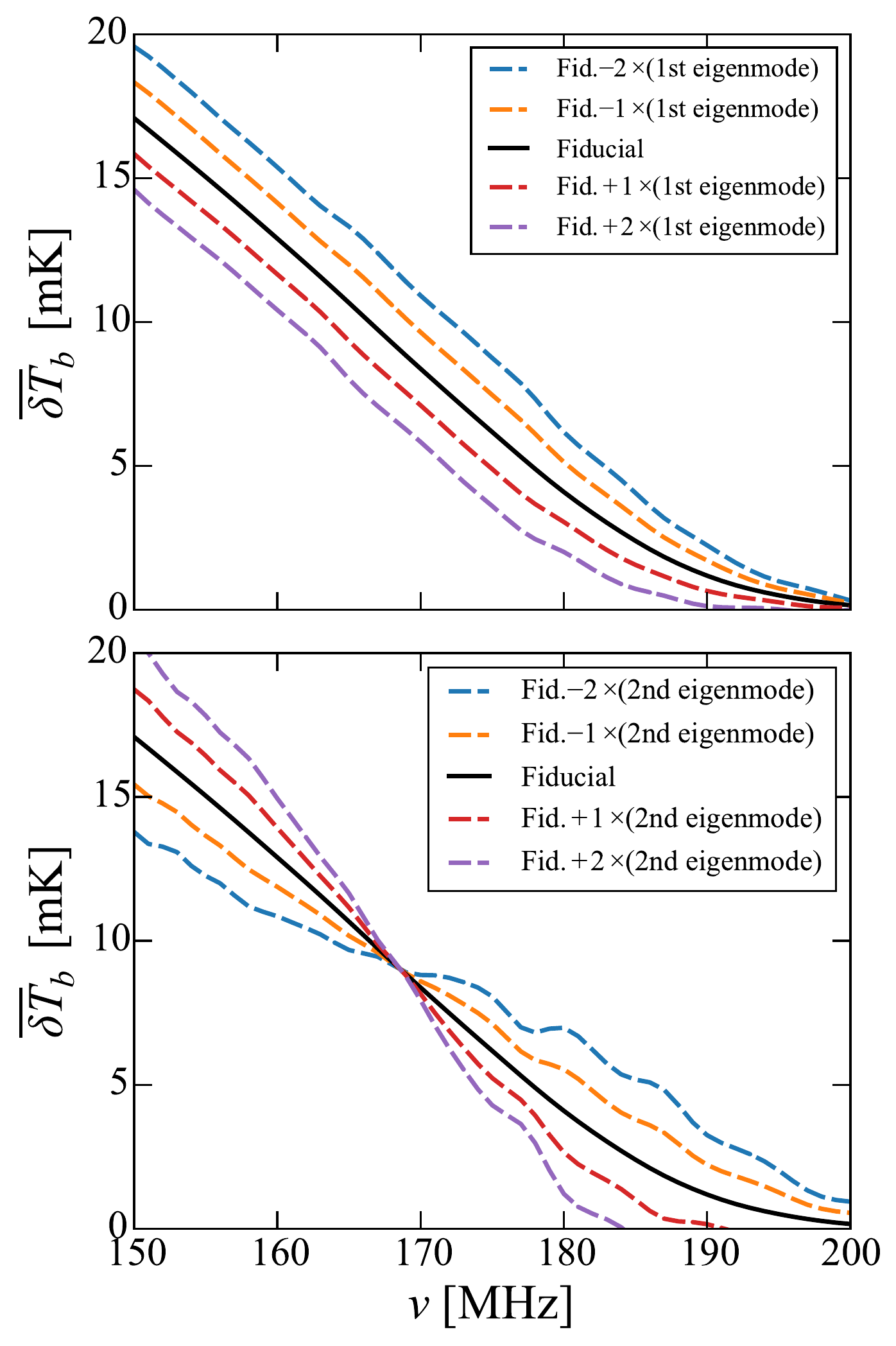}
\caption{Fiducial global signal history (black solid lines) chosen to match the fiducial model tied to the \emph{Planck} TT,TE,EE + lowP + lensing + ext dataset. Dashed lines show perturbations about the fiducial history driven by excitations of the deviation eigenmodes (first eigenmode on top plot; second eigenmode on bottom plot) of our power spectrum-informed principal component basis.}
\label{fig:devEigenmodes}
\end{figure}

To forecast the performance of our fiducial experiment, then, we require an expression for $\Sigma$. We suppose that the data is analyzed using the methods of Ref. \cite{liu_and_parsons2015}. Briefly, we assume that a prior measurement of the $21\,\textrm{cm}$ power spectrum is available, and that these results can be fit to cosmological and astrophysical parameters. Simulations are then run to predict a fiducial global signal history $\overline{\delta T}_b^\textrm{fid} (\nu)$ as well as a plausible set of alternate histories that are allowed within the error bars of the parameters. Forming a covariance matrix of these alternate histories, one may then perform an eigenvalue decomposition to obtain a set of principal component eigenmodes that compactly describe deviations from the fiducial history. The global signal can then be expressed as
\begin{equation}
\overline{\delta T}_b (\nu) = \overline{\delta T}_b^\textrm{\,fid} (\nu) + \sum_i^{N_d} b_i d_i (\nu),
\end{equation}
where $\overline{\delta T}_b^\textrm{\,fid}$ is the fiducial history, $N_d$ is the number of eigenmodes needed to adequately fit the data, $d_i (\nu)$ is the $i$th deviation eigenmode, and $b_i$ its amplitude. The goal of the global signal measurement is to constrain the set of amplitudes $\{ b_i \}$. The effects of changing the amplitudes of the two strongest modes are shown in Fig. \ref{fig:devEigenmodes} for a fiducial history tuned to match \emph{Planck} TT,TE,EE + lowP + lensing + ext. Importantly, we note that even though our deviation eigenmodes are informed by simulations, our global signal measurement remains model-independent, since large deviations from the fiducial global signal history will simply result in stronger measured deviation amplitudes, and possibly a higher $N_d$.

In addition to the deviation mode amplitudes, a global signal experiment must also contend with foreground contamination (in addition to other systematics, which may introduce additional degrees of freedom \cite{switzer_and_liu2014}). Given that the foregrounds are spectrally smooth, we follow previous works \cite{mcquinn_et_al2006} and model them as a sum of $N_p$ Legendre polynomials in $\log \nu$ with a set of foreground amplitudes that are fit alongside the deviation amplitudes. This gives a total of $N_p + N_d$ parameters that are fit for in the analysis of global signal data. To quantify the errors in such fits, we employ the same Fisher matrix formalism that was used in Ref. \cite{presley_et_al2015}, which was in turn based on the treatments of Refs. \cite{pritchard_and_loeb2010,bernardi_et_al2015}. Inverting the Fisher matrix to the obtain a covariance and marginalizing over the nuisance foreground amplitudes, we arrive at an $N_d \times N_d$ matrix $\mathbf{C}$ of error covariances on the deviation amplitudes. These can then be converted into a error covariance matrix $\boldsymbol \Sigma$ between different frequency bins by computing
\begin{equation}
\boldsymbol \Sigma = \mathbf{D}^t \mathbf{C} \mathbf{D},
\end{equation}
where $\mathbf{D}_{ij} = d_i (\nu_j) $. This is essentially the discrete, frequency-space version of $\Sigma ( z, z^\prime)$, the continuous redshift-space covariance that is needed to evaluate Eq. \eqref{eq:DeltaTauDoubleIntegral}. However, the discrete version is sufficient for all intents and purposes, since the deviation eigenmodes can be interpolated and evaluated at whatever frequencies (or redshifts) one desires. With this, Eq. \eqref{eq:DeltaTauDoubleIntegral} can be evaluated to compute the error contribution to $\tau(z)$ from the global signal measurement, which can then be combined in quadrature with the errors from cosmological parameter uncertainties, since the two contributions are independent.

\begin{figure}[!]
\centering
\includegraphics[width=0.5\textwidth]{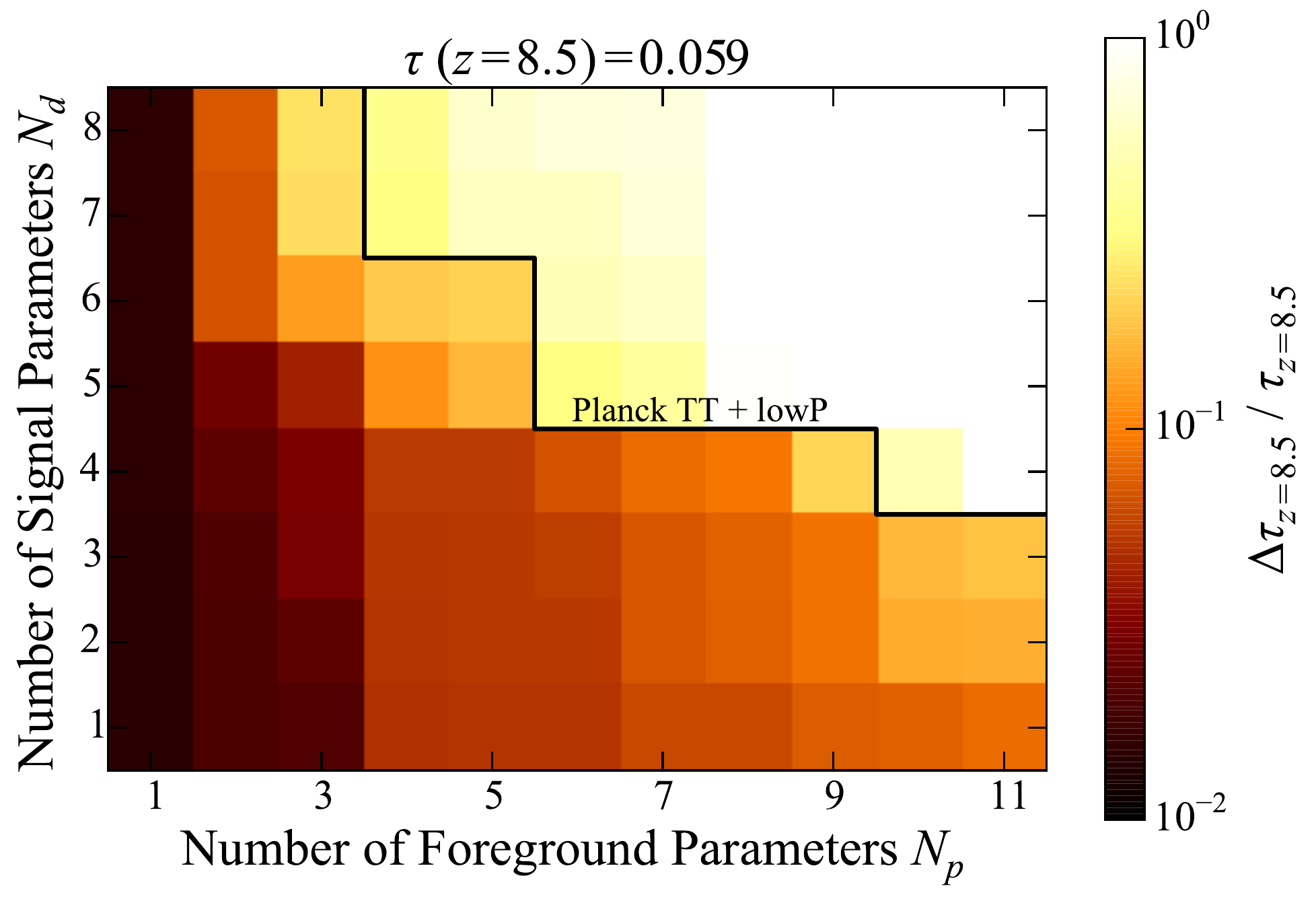}
\caption{Fractional error in a global signal measurement of $\tau(z=8.5)$ as a function of the number of foreground parameters $N_p$ and number of signal parameters $N_d$ that are necessary for an adequate fit to the data. The assumed reionization scenario is chosen to match parameters from \emph{Planck} TT+lowP. A discretized contour of the fractional error on \emph{Planck} measurement of $\tau$ is given by the thick black line for reference. As long as the number of parameters in a global signal remains small, global signal experiments can provide direct, model-independent constraints on relatively low-redshift portions of $\tau$.}
\label{fig:PlanckTTlowP_globalSigErrors}
\end{figure}

\begin{figure}[!]
\centering
\includegraphics[width=0.5\textwidth]{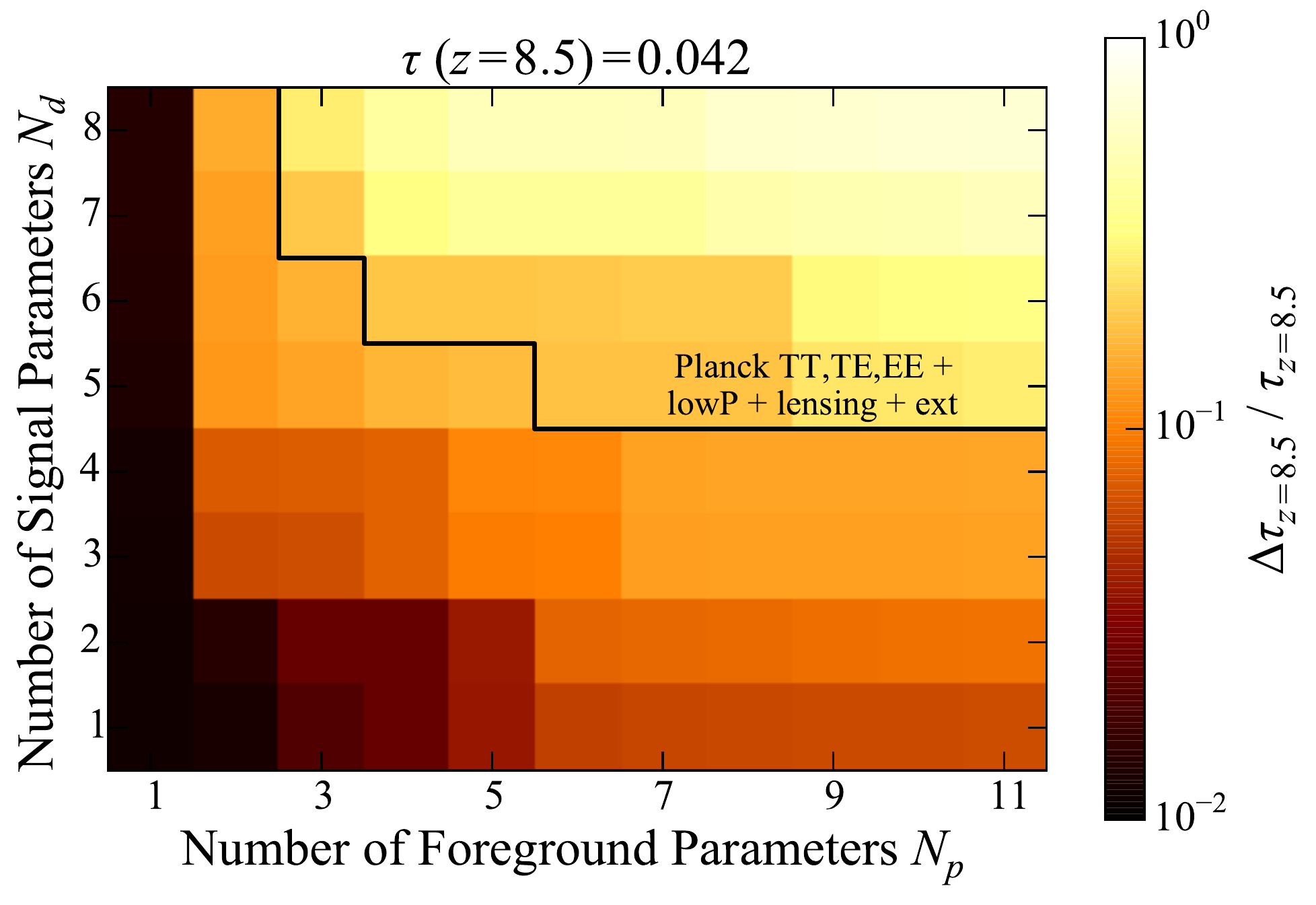}
\caption{Similar to Fig. \ref{fig:PlanckTTlowP_globalSigErrors}, but for the \emph{Planck} TT,TE,EE + lowP + lensing + ext  dataset.}
\label{fig:PlanckTTTEEElowPlensingext_globalSigErrors}
\end{figure}

In Fig. \ref{fig:PlanckTTlowP_globalSigErrors} and Fig. \ref{fig:PlanckTTTEEElowPlensingext_globalSigErrors} we show the forecasts resulting from our analysis for the \emph{Planck} TT+lowP and \emph{Planck} TT,TE,EE + lowP + lensing + ext  datasets, respectively. Shown in color are the fractional errors in a global signal measurement of $\tau (z=8.5)$. One sees that as long as global signal spectra can be fit with relatively few parameters, small error bars on $\tau(z = 8.5)$ can be attained. These typically compare favorably with fractional errors in $\tau$ from \emph{Planck}, which are denoted by the thick black lines on each plot ($\sim 24\%$ for TT+lowP and $\sim 18\%$ for TT,TE,EE + lowP + lensing + ext). We note, however, that these lines are included for reference purposes only and should be interpreted with caution, since any CMB-derived $\tau$ must necessarily be an integral measurement up to the surface of last scattering. From the perspective of a CMB experiment, it is thus impossible to measure $\tau (z=8.5)$ (the contribution to $\tau$ from $0 < z < 8.5$) using the CMB. Nonetheless, the fractional errors from \emph{Planck} convey the rough sense that global signal measurements have the potential to provide independent and competitive constraints on $\tau$.


\section{Conclusions}
\label{sec:conc}

The optical depth parameter $\tau$ serves a dual role in CMB studies. On one hand, it serves as a crude tool for probing reionization, since the optical depth arises from the scattering of CMB photons off free electrons produced during reionization. On the other hand, it can be viewed as a nuisance parameter that simply needs to be marginalized out, in the process degrading the precision of constraints on other cosmological parameters, particularly $A_s$.

In this paper, we advocate the use of highly redshifted $21\,\textrm{cm}$ observations to provide an independent constraint on $\tau$, thereby breaking parameter degeneracies that arise in CMB data analyses (many of which remain even when complementary probes like galaxy surveys are introduced). If modeling uncertainties in our current understanding of reionization can be reduced (as is expected to be case once a cosmological $21\,\textrm{cm}$ detection is made), the opportunity to eliminate $\tau$ as a nuisance parameter has the potential to push CMB observations into a qualitatively new regime, where one would not need to contend with the inherent limitations of solving for $\tau$ internally using CMB data, such as an $\ell_\textrm{min}$ cut-off or eventually, a cosmic variance limit.

We propose two approaches for relating $21\,\textrm{cm}$ observations to $\tau$. Towards the end of the reionization epoch, the complicated astrophysics of the neutral hydrogen spin temperature $T_s$ drops out of the expression for the brightness temperature $\delta T_b$ of the $21\,\textrm{cm}$ line. Measurements of the sky-averaged brightness temperature $\overline{\delta T_b}$ (``global signal measurements") then provide a direct probe of the density-weighted ionized fraction, which can be integrated in redshift to estimate $\tau$. Our forecasts suggest that as along as the observed global signal can be fit without an unreasonably large number of parameters, this technique can be used to provide precise estimates of the lower redshift contributions to $\tau$ (up to, for example, $z \sim 8.5$, but this depends on precisely how reionization proceeds). Provided we limit ourselves to these lower redshift portions of $\tau$
, the resulting global signal estimates of $\tau$ are relatively model-independent and represent an improvement upon the \emph{Planck} constraints.

To compute the full optical depth from $21\,\textrm{cm}$ observations, it is necessary to resort to higher signal-to-noise observations, and here we focus on measurements of the power spectrum $P_{21} (k)$ of $21\,\textrm{cm}$ brightness temperature fluctuations as a function redshift. We envision a scheme where power spectrum measurements are used over a relatively narrow range in redshift (e.g., $6 \le z \le 9$) to constrain reionization parameters. These parameters are then fed into semi-analytic simulations of reionization to predict the density-weighted ionized fraction to high redshifts, which can again be integrated to yield $\tau$. In practice, the simulations themselves depend on cosmological parameters in addition to astrophysical parameters, and to properly account for all uncertainties, parameter estimation must be performed jointly. Under such a scheme, information from the $21\,\textrm{cm}$ line is incorporated by self-consistently requiring the CMB-measured $\tau$ to agree with values of $\tau$ predicted by the $21\,\textrm{cm}$-tuned simulations. Note that we assume that the physics probed by the (relatively narrow) observed redshift range completely describes the physics of $\tau$, i.e., we assume that once the low redshift portions of our model are tuned to observations, the errors in our model extrapolations to higher redshifts are small. Said differently, we assume that any additional physics at higher redshifts makes negligible contributions to $\tau$. These assumptions can be checked using the many examples of proposed or upcoming high redshift probes discussed in Sec. \ref{sec:ModelDependence}.

Forecasting the performance of our method for HERA, we find that while parameter errors are reduced for all cosmological parameters with the introduction of $21\,\textrm{cm}$-derived $\tau$ information, the effects are the most pronounced for $A_s$. This arises because of the known degeneracy between $A_s$ and $\tau$ in CMB observations. With HERA, this degeneracy is broken and errors on $\ln (10^{10} A_s)$ decrease by more than a factor of four. Improved measurements of $A_s$ can sharpen (or alleviate) current tensions between cosmological parameters derived from cluster counts and those from primary CMB anisotropies.

The $21\,\textrm{cm}$ line may also be instrumental in future cosmological detections of the neutrino mass. To obtain precise estimates of the sum of the neutrino masses $\sum m_\nu$, Stage 4 CMB experiments must accurately constrain the low $\ell$ reionization bump signature in their polarization power spectra. Failing to do so would limit the precision of a CMB-derived value of $\tau$, which propagates to a degraded $\sum m_\nu$ constraint since the two parameters are partially degenerate. This degeneracy can be broken by complementing the CMB with $21\,\textrm{cm}$ cosmology. Assuming that multipoles below $\ell_\textrm{min} = 50$ are inaccessible to ground-based Stage 4 experiments, the addition of $21\,\textrm{cm}$ information from HERA improves the $1\sigma$ error bars on $\sum m_\nu$ from $\pm 19\,\textrm{meV}$ to $\pm 12\,\textrm{meV}$. The latter represents a $\sim\!5\sigma$ detection of on the minimum allowed $\sum m_\nu$ of $60\,\textrm{meV}$.

As a first detection and further measurements of the cosmological $21\,\textrm{cm}$ signal are made, it is likely that our understanding of reionization will be considerably refined. We therefore expect our error forecasts to evolve with new data and new models, and our current treatment should be considered only as a proof-of-concept study that demonstrates the potential power of combining $21\,\textrm{cm}$ cosmology with CMB studies. As discussed in Sec. \ref{sec:ModelDependence}, a wide range of near-term and future observations will confirm or refute our models of reionization, with the push to even higher redshifts playing a particularly crucial role in alleviating the potential risks of model-dependent constraints. The present work therefore adds a cosmological motivation to the astrophysical case for pursuing direct, high redshift measurements of the pre-reionization epoch.

Future work can improve upon the results derived in this work by incorporating a greater variety of signatures in the CMB. The essential idea in this paper is to demand self-consistency between reionization constraints from the CMB and the $21\,\textrm{cm}$ line. This requirement of self-consistency need not be limited to $\tau$; our choice to focus on $\tau$ is based simply on the fact that it has a clear degeneracy with $A_s$ and is measured by all CMB experiments. Future high-precision CMB measurements will yield additional constraints on reionization beyond $\tau$, such as those from the kSZ effect. Further details regarding the ionization history may also be obtainable from high-sensitivity polarization measurements \cite{hu_and_holder2003,dai_et_al2015}, with up to five independent modes of the ionization history potentially observable \cite{mortonson_and_hu2008}. All of these constraints can be self-consistently combined with $21\,\textrm{cm}$ measurements and simulations in the manner described in this paper, thus further improving cosmological constraints. Importantly, $21\,\textrm{cm}$ measurements will always remain a crucial check for CMB reionization results, since the CMB does not in general contain enough information to accurately reconstruct the full richness of a physically motivated ionization history \cite{MoradinezhadDizgah_et_al2013}. Additionally, observations of the $21\,\textrm{cm}$ signal may be the first to detect any unexpected features in the ionization history, which may in turn inform how CMB reionization constraints are interpreted.

The aforementioned advances will only serve to improve the already sharp forecasts provided in this work. Upcoming high signal-to-noise measurements of the $21\,\textrm{cm}$ line from arrays such as HERA and SKA will therefore provide not only a transformative understanding of the astrophysics of reionization, but also the opportunity to further push the frontiers of precision cosmology.

\section*{Acknowledgments}
The authors are delighted to acknowledge helpful discussions with James Aguirre, Tzu-Ching Chang, Asantha Cooray, Clive Dickinson, Steve Furlanetto, Nick Gnedin, Bradley Greig, Daniel Jacobs, Andrei Mesinger, Miguel Morales, Michael Mortonson, Peng Oh, Jonathan Pober, Eric Switzer, and Hy Trac. This research was completed as part of the University of California Cosmic Dawn Initiative. AL and ARP acknowledge support from the University of California Office of the President Multicampus Research Programs and Initiatives through award MR-15-328388, as well as from NSF CAREER award No. 1352519, NSF AST grant No.1129258, and NSF AST grant No. 1440343. AL acknowledges support for this work by NASA through Hubble Fellowship grant \#HST-HF2-51363.001-A awarded by the Space Telescope Science Institute, which is operated by the Association of Universities for Research in Astronomy, Inc., for NASA, under contract NAS5-26555. JRP acknowledges funding from the European Research Council under ERC grant \#638743-FIRSTDAWN, from the European Union's Seventh Framework Programme FP7-PEOPLE-2012-CIG grant \#321933-21ALPHA, and from STFC consolidated grant ST/K001051/1. RA is supported by ERC grant \#259505. BDS was supported by a
fellowship from the Miller Institute for Basic Research in Science at the University of California, Berkeley. This research used resources of the National Energy Research Scientific Computing Center, a DOE Office of Science User Facility supported by the Office of Science of the U.S. Department of Energy under Contract No. DE-AC02-05CH11231.

\bibliography{21cmTau}

\end{document}